\documentclass[12pt, draftclsnofoot, onecolumn]{IEEEtran}
 \usepackage{amsmath,amssymb}
 \usepackage{subfigure}
 \usepackage{graphicx,graphics,color,psfrag}
 \usepackage{cite,balance}
 \usepackage{caption}
 \allowdisplaybreaks
 \usepackage{algorithm}
  \usepackage{algorithmic}
 \usepackage{accents}
 \usepackage{amsthm}
 \usepackage{bm}
  \usepackage{url}
 \usepackage[english]{babel}
 \usepackage{multirow}
 \usepackage{enumerate}
 \usepackage{cases}
 \usepackage{stfloats}
 \usepackage{dsfont}
 \usepackage{color,soul}
 \usepackage{amsfonts}
 \usepackage{cite,graphicx,amsmath,amssymb}
 \usepackage{subfigure}
 \usepackage{fancyhdr}
 \usepackage{hhline}
 \usepackage{graphicx,graphics}
 \usepackage{array,color}
 \usepackage{amsmath}

\newtheorem{lemma}{\emph{\underline{Lemma}}}

\def\phi{\varphi}

\def\l{\left}
\def\r{\right}
\def\({\left(}
\def\){\right)}

\def\b0{{\mathbf{0}}}

\newcommand{\nn}{\nonumber}

\begin{document}
\captionsetup[figure]{name={Fig.}}
\title{\huge 
%\vspace{-10pt}
%\title{\LARGE 
3D Trajectory Optimization in Rician Fading\\ for UAV-Enabled Data Harvesting}
\author{Changsheng You and Rui Zhang   \thanks{\noindent C. You and R. Zhang are  with the Dept. of Electrical and Computer Engineering, National University of Singapore, Singapore (Email: eleyouc@nus.edu.sg, elezhang@nus.edu.sg).
 }}

\maketitle
\vspace{-10pt}
\begin{abstract}
Dispatching unmanned aerial vehicles (UAVs) to harvest sensing-data from distributed sensors is expected to significantly improve the data collection efficiency in conventional wireless sensor networks (WSNs). In this paper, we consider a UAV-enabled WSN where  a flying UAV is employed to collect data from multiple sensor nodes (SNs). Our objective is to maximize the minimum average data collection rate from all SNs subject to a prescribed reliability constraint for each SN by jointly optimizing the UAV communication scheduling and \emph{three-dimensional} (3D) trajectory. Different from the existing works that assume the simplified line-of-sight (LoS) UAV-ground channels, we consider the more practically accurate \emph{angle-dependent} \emph{Rician fading} channels between the UAV and SNs  with the Rician factors determined by the corresponding UAV-SN elevation angles. However, the formulated optimization problem is intractable due to the lack of a closed-form expression for a key parameter termed  \emph{effective fading power} that characterizes the achievable rate given the reliability  requirement in terms of outage probability. To tackle this difficulty, we first approximate the parameter by a \emph{logistic} (`S' shape) function  with respect to the 3D UAV  trajectory by using the data regression method. Then the original problem is reformulated to an approximate form, which, however, is still challenging to solve due to its non-convexity.  As such, we further propose an efficient algorithm to derive its suboptimal solution by using the block coordinate descent technique, which iteratively optimizes the communication scheduling, the UAV's horizontal trajectory, and its vertical trajectory. The latter two subproblems are shown to be non-convex, while locally optimal solutions are obtained for them by using the successive convex approximation technique. Last, extensive numerical results are provided to evaluate the performance of the proposed algorithm and draw new insights on the 3D UAV trajectory  under the Rician fading as compared to conventional  LoS channel models.
\end{abstract}
\vspace{-10pt}

\section{Introduction}
Unmanned aerial vehicles (UAVs) (or Drones) are expected to be widely deployed in the future  for enabling a proliferation of applications ranging from aerial delivery to surveillance and monitoring, disaster rescue, and remote sensing\cite{zeng2016wireless,baek2018design,liu2010dynamic}. Among others, dispatching UAVs to harvest sensing-data from distributed sensor nodes (SNs) is anticipated  to be a promising technology for realizing the future  Internet of Things (IoT). Different from conventional  wireless sensor networks (WSNs) that rely on static data collecting nodes and multihop data relaying among the SNs, the UAV-enabled WSN leverages a mobile data collector  mounted on the UAV that communicates with the SNs directly by exploiting the line-of-sight (LoS) dominant UAV-ground channels.  This helps not only significantly improve the WSN coverage and throughput, but also effectively reduce energy consumption of the SNs by scheduling their transmissions based on the UAV's trajectory. Such advantages have attracted growing research attention in recent years on UAV-enabled WSNs, including  the designs of UAV trajectory, SN wakeup schedule, trajectory-aware signal modulation and coding, etc \cite{abdulla2014optimal,liu2018energy,liu2018age,zhan2018energy,zhan2018trajectory,gong2018flight,ebrahimi2018uav,zeng2018trajectory,yang2018energy}. In particular, for the trajectory design in UAV-enabled WSNs, most of the existing works (e.g., \cite{liu2018age,zhan2018energy,zhan2018trajectory,gong2018flight,ebrahimi2018uav,zeng2018trajectory,yang2018energy}) assumed that the UAV flies at a fixed (minimum) altitude and thus only the  two-dimensional (2D) UAV trajectory was considered. In contrast, our current work further exploits the vertical trajectory of the UAV and presents a new design framework of \emph{three-dimensional} (3D) UAV trajectory to improve the rate performance in UAV-enabled WSNs.

Besides UAV-enabled WSNs, UAV trajectory design has been widely investigated in other wireless networks, such as UAV-assisted terrestrial communication systems \cite{zeng2017energy,wu2018joint,wu2018uav,wu2018common}, relay systems \cite{zeng2016throughput}, cellular networks \cite{zhang2018cellular}, radio access networks \cite{zhang2018uav}, and  wireless power transfer networks \cite{xu2018uav}.
The UAV trajectory design critically  relies on the UAV-ground channel modelling. Among others, there are three commonly adopted channel models in the literature, including the  LoS channel, probabilistic LoS channel, and Rician fading channel. Particularly, as the UAV at a sufficiently high altitude has a high likelihood to establish an LoS link with the ground node \cite{3GPPUAV}, the deterministic LoS channel following the free-space pathloss model has been widely used in most of the existing works on UAV trajectory design (e.g., \cite{liu2018age,zhan2018energy,zhan2018trajectory,gong2018flight,ebrahimi2018uav,zeng2018trajectory,yang2018energy}) due to its convenience for optimization. However, such a simplified model may be practically inaccurate in urban/suburban areas, as it neglects the stochastic shadowing and small-scale fading. Considering shadowing, the signal propagation can be blocked by obstacles (e.g., buildings) in urban areas and thereby the UAV-ground channels can be largely categorized  into either LoS or non-LoS (NLoS) link at different locations with different characteristics. To avoid the excessive measurement cost for attaining the complete information of LoS/NLoS channels at all locations in a large geographical area, a statistical probability based model  for the occurrence of LoS/NLoS channels was proposed in \cite{holis2008elevation} as a logistic function, whose parameters are determined by the specific environment and the elevation angle of the UAV. Based on this empirical model, substantial  research has been conducted for designing the 3D UAV placement  to optimize  the communication performance in terms of coverage, throughput, delay, and reliability (see e.g., \cite{lyu2017placement,mozaffari2016unmanned,mozaffari2016efficient,mozaffari2017mobile,bor2016efficient,alzenad2018} and the references therein). This channel model, although being suitable for communication performance analysis, cannot be directly applied to design UAV trajectory. The reason is that along the UAV trajectory, the LoS probability in a local region generally is not identical to that averaged over the whole area of interest, and it is also spatially correlated depending on the surrounding environment. The work \cite{esrafilian2018learning} made the first attempt to tackle this difficulty by learning the local channel parameters and constructing a local 3D radio map, based on which the UAV trajectory was designed by a novel map compression method.  
 Another widely adopted model is the Rician fading model that comprises a deterministic LoS component and a random multipath component due to reflection, scattering, and diffraction by the ground obstacles \cite{khuwaja2018survey}. This model is suitable for urban/suburban areas with the UAV at a sufficiently high altitude with less shadowing but non-negligible small-scale fading. The Rician factor, as reported in \cite{matolak2017air}, is affected by the communication band (L/C band), surrounding environment, and the UAV-ground elevation angle. As the elevation angle enlarges, the experimental results in \cite{shimamoto2006channel} show that the Rician factor tends to \emph{exponentially} increase since a larger elevation angle is likely to incur less ground reflection, scattering, and obstruction. Such an (elevation) \emph{angle-dependent} Rician fading model is more practically accurate than the conventional simplified LoS model, but the UAV trajectory design under this model has not yet been investigated in the existing literature. This thus motivates our current work as the first attempt to design the 3D UAV trajectory in the angle-dependent Rician fading channel.

Specifically, in this paper, we consider a UAV-enabled WSN where a UAV flies over multiple SNs to collect data from them. The SNs are normally in the silent mode for energy saving and transmit data only when being waken up by the UAV (e.g., by broadcasting a beacon signal) and scheduled for transmission.  Our objective is to maximize the minimum average data collection rate from all SNs,  while ensuring that the sent data is reliably received by the UAV with an outage probability less than a prescribed value.  Compared with the existing works on UAV trajectory design relying on the simplified LoS channel model, adopting the angle-dependent Rician fading channel model introduces new design issues. First, as the UAV-SN channel is not fully predictable along the UAV trajectory due to the random small-scale fading, we should consider an \emph{outage-aware adaptive-rate}  transmission scheme. However, the relationship between the resultant  achievable rate and the UAV trajectory is \emph{non-trivial}  due to the distance-dependent pathloss and angle-dependent Rician factor. Second, unlike the conventional 2D trajectory designs with a fixed (minimum) UAV altitude, the elevation angle-dependent Rician fading calls for a joint optimization of both the horizontal and vertical UAV trajectories, leading to the said 3D UAV trajectory design.  Tackling these key issues yields the main contributions of this paper as summarized  below.
\begin{itemize}
\item First, we formulate an optimization problem to maximize the minimum data collection rate from all SNs with an outage probability guarantee for each SN by jointly designing the UAV communication scheduling and 3D trajectory. This problem, however, is intractable due to the lack of a closed-form expression for a  key parameter termed \emph{effective fading power}, which characterizes the achievable rate in the fading channel under the given outage probability constraint.  To address this difficulty, by leveraging the data regression method, we approximate the effective fading power by a logistic function with respect to (w.r.t.) the 3D UAV trajectory, and then reformulate the original problem into a tractable form.
 \item Next, we propose an efficient algorithm for solving the reformulated non-convex problem. Specifically, we first apply continuous relaxation to the integer UAV scheduling  constraint and then solve  the relaxed  problem by using the block coordinate descent (BCD) technique that iteratively optimizes the UAV communication scheduling, horizontal trajectory, and vertical trajectory. Since the  subproblems for optimizing the UAV horizontal and vertical trajectories are still non-convex,  the successive convex approximation (SCA) technique is applied to derive the locally optimal solutions to them.
 \item Last, numerical results are provided to evaluate the performance of the proposed algorithm and compare the optimized UAV trajectories under the considered Rician fading and conventional LoS channel models. We  show that for the case with one single SN, the proposed UAV trajectory can exploit the vertical trajectory (altitude variation) to enhance the data collection rate, especially given a high maximum vertical speed and a stringent outage probability requirement. Moreover, the designed UAV trajectory is close to that assuming the LoS channel when the Rician fading approaches to either Rayleigh fading or LoS channel. Furthermore, for the case with multiple SNs, by leveraging the \emph{angle-aware} joint horizontal and vertical trajectory design, the proposed 3D UAV trajectory  can significantly enhance the performance over the conventional trajectory assuming the simplified LoS channel.
 \end{itemize}

The remainder  of this paper is organized as follows. Section~\ref{Sec:Model}
presents the system model and problem formulation. In Section \ref{Sec:FunAppro}, the effective-fading-power function is approximated by a logistic form, based on which the optimization problem is reformulated. Subsequently,  we  propose an efficient algorithm for solving the reformulated problem in Section~\ref{Sec:PropAlgo}. Numerical results are provided in Section~\ref{Sec:Nume}, followed by the conclusions in Section~\ref{Sec:Conc}.

%\vspace{-0.5mm}
\section{System Model and Problem Formulation}\label{Sec:Model}

\begin{figure}[t]
\begin{center}
\includegraphics[height=6.6cm]{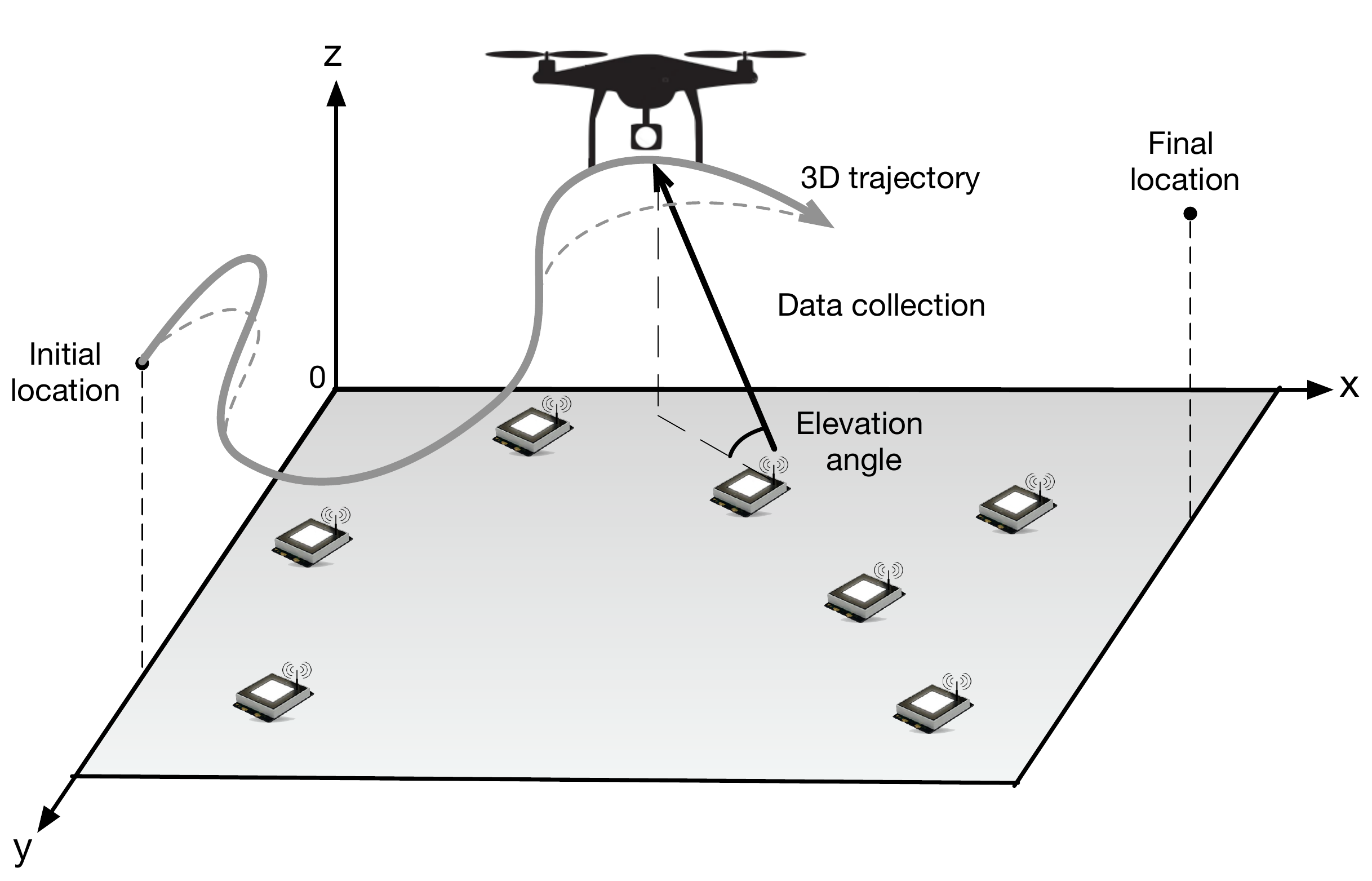}
\caption{UAV-enabled data collection.}
\label{Fig:System}
\end{center}
\end{figure}

Consider a UAV-enabled WSN as shown in Fig.~\ref{Fig:System} with a UAV flying over $N$ ground SNs and collecting data from them within a given time duration of $T$. The SNs are indexed by the set $\mathcal{N}=\{1,\cdots, N\}$ and their individual location is represented by the 3D Cartesian coordinate $({\mathbf{w}}_n^{T}, 0)$ for $n\in \mathcal{N}$, with ${\mathbf{w}}_n=[x_n, y_n]^{T}\in\mathbb{R}^{2\times1}$ denoting the horizontal coordinate. The UAV's  initial and final  locations are predetermined, represented by $({\mathbf{q}}_0^{T}, z_0)$  and  $({\mathbf{q}}_{F}^{T}, z_F)$, respectively, where ${\mathbf{q}}_0^{T}\in\mathbb{R}^{1\times2}$ and ${\mathbf{q}}_{F}^{T}\in\mathbb{R}^{1\times2}$ denote the horizontal coordinates, and $z_0$ and $z_F$ are the corresponding altitudes. Assuming that the UAV has prior information of all SNs' locations, we jointly design the UAV communication scheduling and 3D trajectory  for maximizing the minimum average data collection rate from all SNs, under the constraints on the UAV communication scheduling and 3D trajectory, while ensuring data being reliably received by the UAV	 under a given tolerable
 outage probability.

\vspace{-2mm}
\subsection{UAV Trajectory Model}

For ease of exposition,  the time horizon $T$ is discretized into $M$ equal time slots, indexed by $\mathcal{M}=\{1,\cdots, M\}$. The elemental slot length $\delta=T/M$ is appropriately chosen such that the UAV's location can be assumed to be approximately unchanged within each time slot. Then the UAV trajectory can be approximated by the $(M+1)$-length 3D  sequence $\{({\mathbf{q}}[m]^T, z[m])\}$, with ${\mathbf{q}}[m]=[x[m], y[m]]^{T}$ and $z[m]$ respectively denoting the horizontal  and vertical coordinates. Assume that the UAV can independently control the horizontal and vertical flying speeds with the maximum speeds denoted by $V_{\rm{xy}}$ and $V_{\rm{z}}$ in meter/second (m/s), respectively \cite{sun2018optimal}. Then the maximum horizontal and vertical flying distances within each time slot are $S_{\rm{xy}}=V_{\rm{xy}}\delta$ and $S_{\rm{z}}=V_{\rm{z}}\delta$, respectively, leading to the following UAV flying speed constraints:
\begin{equation*}
\vspace{-0.5mm}
||{\mathbf q}[m+1]-{\mathbf q}[m]||\le S_{\rm{xy}} ~~\text{and}~~ |z[m+1]-z[m]|\le S_{\rm{z}}, \quad \forall m\in \mathcal{M}.
\end{equation*}
In addition, the predetermined initial and final locations for the UAV enforce:
\begin{equation*}
\vspace{-0.5mm}
({\mathbf q}[1]^T, z[1])=({\mathbf{q}}_0^{T}, z_0) ~~\text{and}~~({\mathbf q}[M+1]^T, z[M+1])=({\mathbf{q}}_F^{T}, z_F).
\end{equation*} 
Last, to avoid obstacles and maintain the LoS paths between the UAV and SNs, the UAV is required to fly above the SNs with a minimum altitude $H$, leading to: $z[m]\ge H, \forall m$.
\vspace{-1mm}
\subsection{UAV-Ground Channel Model}
 
We consider the block fading channels for the UAV-SN links, where the channel remains unchanged within each fading block and independently changes over different blocks. As the duration of each fading block is typically much smaller than that of each time slot,  we assume each time slot consisting of $L>1$ fading blocks. As reported in practical experiments, the UAV at a sufficiently high altitude is likely to establish LoS links with the ground SNs and also experiences small-scale fading due to rich scattering  \cite{khawaja2018survey}. Therefore, the channel between each SN, say SN $n$, and the UAV in the $\ell$-th fading block of time slot $m$ can be modeled as 
\begin{equation}\label{Eq:Channel}
h_n[m,\ell]=\sqrt{\beta_n[m]} g_n[m,\ell],
\end{equation} where $\beta_n[m]$ is the large-scale average channel power gain accounting for signal attenuation including pathloss and shadowing
% \footnote{\color{black} In general, the large-scale channel power gain includes the pathloss and shadowing which are also effected by the elevation angle, but is fixed in the current work for tractable analysis.} 
 and $g_n[m,\ell]$ is the small-scale fading coefficient. Specifically, let $d_n[m]$ denote the the distance between the UAV and SN $n$ in time slot $m$, given by
\begin{equation}
d_n[m]=\sqrt{|| {\mathbf{q}}[m]-{\mathbf{w}}_n||^2+z[m]^2}. \label{Eq:Distance}
\end{equation}
Then, the average channel power gain, $\beta_n[m]$, can be modeled as 
%  the large-scale fading includes both effects of pathloss and shadowing, and  the average channel power gain, $\beta_n[m]$,can be   follows the free-space pathloss model and thereby the average channel power gain, $\beta_n[m]$, can be 
%modeled as
\begin{equation}
{\color{black}\beta_n[m]=\beta_0 d_n^{-\alpha}[m],}
%=\beta_0/(|| {\mathbf{q}}[m]-{\mathbf{w}}_n||^2+z[m]^2)
\label{Eq:Pathloss}
\end{equation}
where $\beta_0$ is the average channel power gain at the reference distance $d_0=1$ m, {\color{black}{$\alpha$ is the pathloss exponent that usually has a value between $2$ and $6$.}}\footnote{\color{black} In general, for UAV communications, the pathloss exponent and variance of random shadowing are also functions of the elevation angle between the UAV and SN, but are assumed constant in the current work for tractable analysis.}
%\footnote{\color{black}{This work can be extended to other pathloss exponents in practice with the main solution approach remaining largely unchanged.}} 
%\begin{equation}
%\beta_n[m]=\beta_0 d_n^{-\alpha}[m],
%%=\beta_0/(|| {\mathbf{q}}[m]-{\mathbf{w}}_n||^2+z[m]^2)
%\label{Eq:Pathloss}
%\end{equation}
%where $\beta_0$ is the average channel power gain at the reference distance $d_0=1$ m, $\alpha\ge2$ is the pathloss exponent, and \begin{equation}
%d_n[m]=\sqrt{|| {\mathbf{q}}[m]-{\mathbf{w}}_n||^2+z[m]^2} \label{Eq:Distance}
%\end{equation} is the distance between the UAV and SN $n$ in time slot $m$. 
Next, due to the existence of the LoS path, the small-scale fading can be modeled by the Rician fading below  with $\mathbb{E}[|g[m,\ell]|^2]=1$:
\begin{equation}
g_n[m,\ell]=\sqrt{\frac{\tilde{K}_n[m,\ell]}{\tilde{K}_n[m,\ell]+1}} g+\sqrt{\frac{1}{\tilde{K}_n[m,\ell]+1}} \tilde{g}, \label{Eq:SmallFading}
\end{equation}
where $g$ denotes the deterministic LoS channel component with $| g |=1$, $\tilde{g}$ represents the random scattered component which is a zero-mean unit-variance circularly symmetric complex Gaussian (CSCG) random variable, and $\tilde{K}_n[m,\ell]$ denotes  the Rician factor of the channel between SN $n$ and the UAV in the $\ell$-th fading block of time slot $m$. Note that due to the mobility of the UAV, the Rician factors for SN $n$ in different time slots are in general \emph{non-identical}, which is closely related to the elevation angle  between the SN and the UAV (see Fig.~\ref{Fig:System}) as reported in prior  experiments \cite{shimamoto2006channel}.  Particularly, as the elevation angle increases, the UAV-SN link tends to experience less scattering and thus  includes a larger portion of LoS component, leading to an increasing Rician factor. Since  the elevation angle within each time slot has negligible change, the Rician factors in different fading blocks over the same time slot  are assumed \emph{identical}, i.e., $\tilde{K}_n[m,\ell] = K_n[m], \forall \ell$. In other words, the channels in the same time slot are identically distributed. Based on the experimental results in \cite{shimamoto2006channel}, the angle-dependent Rician factor can be modeled by the following exponential function:
\begin{equation}
K_n[m]=A_1\exp(A_2\theta_n[m]), \label{Eq:Kfactor}
\end{equation}
where $\theta_n[m]$ is the elevation angle given by 
\begin{equation}
\theta_n[m]=\arcsin(z[m]/d_n[m]),\label{Eq:angle}
\end{equation} $A_1$ and $A_2$ are constant coefficients depending on the specific environment. Then we have $K_{\min}\le K_n[m]\le K_{\max}$, where $K_{\min}=A_1$ and $K_{\max}=A_1e^{A_2\pi/2}$. It is worth noting that, for each SN, the distributions of the small-scale fading in different time slots are \emph{correlated} and determined by the 3D UAV trajectory, which can be observed from \eqref{Eq:SmallFading}--\eqref{Eq:angle}, making the optimization problem formulated in the sequel highly challenging to solve.

\vspace{-1mm}

\subsection{Data Collection Model}
Assume that each SN transmits data with a constant transmission power $P_n$ only when being waken up by the UAV and scheduled for transmission, and otherwise keeps in the silent mode for energy saving.  Let $a_n[m]$ denote the binary UAV communication scheduling for SN $n$ in time slot $m$, where the SN wakes up if $a_n[m]=1$ and sleeps otherwise. Assume that only one SN is scheduled for  transmission in each time slot, leading to the following  scheduling constraints:
\begin{align*}
\sum_{n=1}^{N} a_n[m]\le 1, \quad \forall m, \quad\text{and}\quad a_n[m]\in\{0,1\}, \quad \forall n,m.
\end{align*}

At the beginning, the UAV determines its trajectory $\{{\mathbf q}[m], z[m]\}$ and communication scheduling $\{a_n[m]\}$ using the knowledge of SNs' locations,  which is assumed to be known at the UAV. Then along the flying trajectory, the UAV wakes up the corresponding SN in each time slot and informs it the transmission rate via the downlink reliable control channel. Consider data transmission for each SN, say SN $n$. In each time slot $m$, if the SN is waken up, the maximum achievable rate between the SN and the UAV during the $\ell$-th fading block of time slot $m$, denoted by $C_n[m,\ell]$ in bits/second/Hertz (bps/Hz), is given as 
\begin{equation}
C_n[m,\ell]=\log_2\l(1+\dfrac{|h_n[m,\ell]|^2 P_n}{\sigma^2 \Gamma}\r),
\end{equation} 
where $\sigma^2$ is the receive noise power and $\Gamma>1$ denotes the signal-to-noise ratio (SNR) gap between the practical modulation-and-coding  scheme and the theoretical Gaussian signaling. However, for UAV trajectory design, the above rate is not exactly known due to the lack of the knowledge for instantaneous channels (i.e., $\{h_n[m,\ell]\}$) prior to the UAV's flight.  Since the SN-UAV channel is independent and identically distributed (i.i.d.) in the fading blocks of the same time slot and non-identically distributed in different time slots, we consider the adaptive-rate transmission scheme at the SNs. To be specific, each SN transmits data at a fixed  rate $R_n[m]$ in the fading blocks of the same time slot and different  rates over different time slots. Therefore, the outage probability that the UAV cannot successfully receive the transmitted data from SN $n$ in the $\ell$-th fading block of time slot $m$ can be expressed as
\begin{align}
p_n[m,\ell]&=\mathbb{P}(C_n[m,\ell]<R_n[m]) \nn \\
&=\mathbb{P}\l(|g_n[m,\ell]|^2<\frac{\sigma^2 \Gamma(2^{R_n[m]}-1)}{\beta_n[m] P_n}\r) \nn\\
&=F_{n,m}\l(\frac{\sigma^2 \Gamma(2^{R_n[m]}-1)}{\beta_n[m] P_n}\r),
\label{Eq:OutProb}
\end{align}
where $F_{n,m}(u)$ denotes the cumulative distribution function (cdf) of the random variable $|g_n[m,\ell]|^2$ which  is a non-decreasing function w.r.t. $R_n[m]$. For Rician fading, the cdf of $|g_n[m,\ell]|^2$ 
can be explicitly expressed as
\begin{equation}
F_{n,m}(u)=1-Q_1\l(\sqrt{2K_n[m]}, \sqrt{2(K_n[m]+1)u}\r),\label{Eq:cdf}
\end{equation}
where $Q_1(x, y)$ is the standard Marcum-Q function. Note that in the same time slot,  the outage probability of SN $n$, $p_n[m,\ell]$, is identical over different fading blocks and thus is re-denoted by $p_n^{\rm{out}}[m]$. To maximize the data collection rate and ensure the transmitted data from all SNs in different fading blocks being reliably received by the UAV, $R_n[m]$ is chosen such that $p_n^{\rm{out}}[m]=\epsilon, \forall n, m$, where $\epsilon$ is the maximum tolerable outage probability which is typically in the range of $0<\epsilon\le0.1$ in practice. Combining  \eqref{Eq:OutProb} and \eqref{Eq:Pathloss} with $p_n[m,\ell]=\epsilon$ yields the \emph{outage-aware} achievable rate $R_n[m]$ given by:
\begin{equation}
{\color{black}R_n[m]=\log_2\l(1+\frac{f_{n}[m] \gamma_n }{(|| \mathbf{q}[m]-\mathbf{w}_n||^2+z[m]^2)^{\alpha/2}}\),}\label{Eq:Rate}
\end{equation}
where  $\gamma_n\overset{\triangle}{=}\frac{P_n \beta_0}{\sigma^2\Gamma}$ and $f_{n}[m]$ denotes the unique solution to $F_{n,m}(u)=\epsilon$. When $F_{n,m}(1)<\epsilon$, we set $f_n[m]=1$. It is worth noting that given a fixed maximum tolerable outage probability, $f_{n}[m]$ is determined by the cdf $F_{n,m}(u)$, which in turn implicitly depends on the 3D UAV trajectory $\{\mathbf{q}[m], z[m]\}$ via the Rician factor $K_n[m]$ and the elevation angle $\theta_n[m]$, as shown in \eqref{Eq:Distance}--\eqref{Eq:angle} and \eqref{Eq:OutProb}--\eqref{Eq:cdf}. As a result, $f_{n}[m]$ can be equivalently denoted as a function of the UAV trajectory: $f_n[m]=\phi_n(\mathbf{q}[m], z[m])$ where the function $\phi_n(x,y)$, however,  has no explicit form and will be approximated in the sequel. Intuitively, $f_n[m]$ can be understood as the effective fading power that guarantees reliable transmission of SN $n$ given the outage probability requirement leading to the achievable rate  given in \eqref{Eq:Rate},  and thus is termed as \emph{effective fading power}. As such, $\phi_n(\mathbf{q}[m], z[m])$ is called the \emph{effective-fading-power function}.

\subsection{Problem Formulation}
Our objective is to maximize the minimum average data collection rate from all SNs under the constraints on the  UAV communication scheduling and 3D trajectory, while ensuring data being reliably collected given the maximum tolerable outage probability. Based on the preceding models, this problem can be formulated as follows.
\begin{subequations}
\begin{align}({\bf P1})
\max_{\l\{\substack{\mathbf{q}[m], z[m], \\ a_n[m]}\r\}, \eta } ~~&\eta \nn \\  
\text{s.t.}~~&{\color{black}\frac{1}{M}\sum_{m=1}^{M} a_n[m]\log_2\l(1+\frac{\phi_n(\mathbf{q}[m], z[m]) \gamma_n }{(|| \mathbf{q}[m]-\mathbf{w}_n||^2+z[m]^2)^{\alpha/2}}\)}\ge \eta, \forall n\in \mathcal{N}, \label{Eq:MinRateCons}\\
&||{\mathbf q}[m+1]-{\mathbf q}[m]||\le S_{\rm{xy}},  \qquad\qquad\qquad  \forall m\in \mathcal{M}, \label{Eq:P1ConsStar}\\
& |z[m+1]-z[m]|\le S_{\rm{z}}, \quad\qquad\qquad\qquad~\forall m\in \mathcal{M}, \label{Eq:P1VerSpeed}\\
& ({\mathbf q}[1]^T,z[1])=({\mathbf{q}}_0^{T}, z_0), ~~~({\mathbf q}[M+1]^T, z[M+1])=({\mathbf{q}}_F^{T}, z_F),\\
& z[m]\ge H, \quad\qquad\qquad\qquad\qquad\qquad\qquad\forall m\in \mathcal{M},\label{Eq:zmin}\\
&\sum_{n=1}^{N} a_n[m]\le 1, \quad\qquad\qquad\qquad\qquad\qquad\forall m\in \mathcal{M}, \label{Eq:ScheCons}\\
&a_n[m]\in\{0,1\}, \qquad\qquad\qquad\qquad\qquad\quad \forall n\in \mathcal{N},m\in \mathcal{M}. \label{Eq:P1InterCons}
\end{align}
\end{subequations}

Note that the derivation for the optimal solution to Problem P1 is \emph{intractable} due to the lack of a closed-form expression for the effective-fading-power function, $\phi_n(\mathbf{q}[m], z[m])$,  which can be observed from \eqref{Eq:cdf} since $f_{n}[m]=\phi_n(\mathbf{q}[m], z[m])$ is the inverse of the standard Marcum-Q function whose exact value can only be computed by iterative algorithms (e.g., \cite{gil2014asymptotic}). As a result, the rate constraint \eqref{Eq:MinRateCons}  relates to the 3D UAV trajectory variables, $\{\mathbf{q}[m], z[m]\}$, in an implicit  manner. Moreover, the UAV scheduling variables, $\{a_n[m]\}$, are binary and hence incur the integer constraint \eqref{Eq:P1InterCons}. To address these issues, we propose an efficient algorithm to derive the suboptimal solution to Problem P1. The key idea is to firstly reformulate the optimization problem by approximating the effective-fading-power function  in a tractable form,  and then design efficient UAV scheduling and 3D trajectory by solving the reformulated problem, which are elaborated in the following  sections.

\section{Approximation for effective fading power and Problem Reformulation}\label{Sec:FunAppro}
In this section, we approximate the effective-fading-power function by using the logistic regression method, based on which the optimization problem is reformulated. 
\subsection{Approximation for Effective Fading Power}
The intractability for the effective-fading-power function is due to the lack of an explicit form for the inverse Marcum-Q function. This issue can be addressed by approximating the inverse Marcum-Q function as derived in \cite{azari2018ultra}. Combining it with  \eqref{Eq:cdf} yields the result given in the following lemma. In this subsection, the superscripts of  notations are omitted for ease of exposition without incurring confusion.

\begin{lemma}\label{Lem:FFAppro1}\emph{Considering practical outage probability requirement (e.g., $0<\epsilon\le0.1$), the effective fading power $f$ can be approximated in the following closed form by using the approximation for the inverse Marcum-Q function in \cite{azari2018ultra}:
\begin{equation}
f\approx\bar{f}\overset{\triangle}{=}\frac{w^2}{2(K+1)},\label{Eq:GAppro}
\end{equation}
where  $w$ is defined by
\begin{equation}
w=\begin{cases}
  \sqrt{-2\ln(1-\epsilon)} e^{\frac{K}{2}},&\mbox{$K\le \frac{K^2_{\rm{th}}}{2}$},\\
   \sqrt{2K}+\frac{1}{2Q^{-1}(\epsilon)} \ln\l(\frac{\sqrt{2K}}{\sqrt{2K}-Q^{-1}(\epsilon)}\r)-Q^{-1}(\epsilon),  &\mbox{$K> \frac{K^2_{\rm{th}}}{2}$},
   \end{cases}\label{Eq:U_n}
\end{equation}
$K_{\rm{th}}$ is the intersection of the sub-functions at $\sqrt{2K}>Q^{-1}(\epsilon)$, and $Q^{-1}(x)$ is the inverse Q-function.}
\end{lemma}

The approximation in Lemma~\ref{Lem:FFAppro1} can be shown to be largely accurate by numerical results (omitted for brevity). Nevertheless, although Lemma~\ref{Lem:FFAppro1} approximates the effective fading power $f$ in an explicit form,  it still relates with the 3D UAV trajectory $\{\mathbf{q}, z\}$ in a complicated manner via the Rician factor and elevation angle, which can be observed from \eqref{Eq:GAppro}, \eqref{Eq:U_n},  and \eqref{Eq:Distance}--\eqref{Eq:angle}.  This makes it hard to characterize the effects of 3D UAV trajectory on the effective fading power and hence on the outage-aware achievable rate, and in turn renders the optimization problem still challenging to solve. 

\begin{figure}[t!]
\centering
\subfigure[Maximum tolerable outage probability $=0.01$.]{\label{Fig:Regression3}
\includegraphics[height=5.3cm]{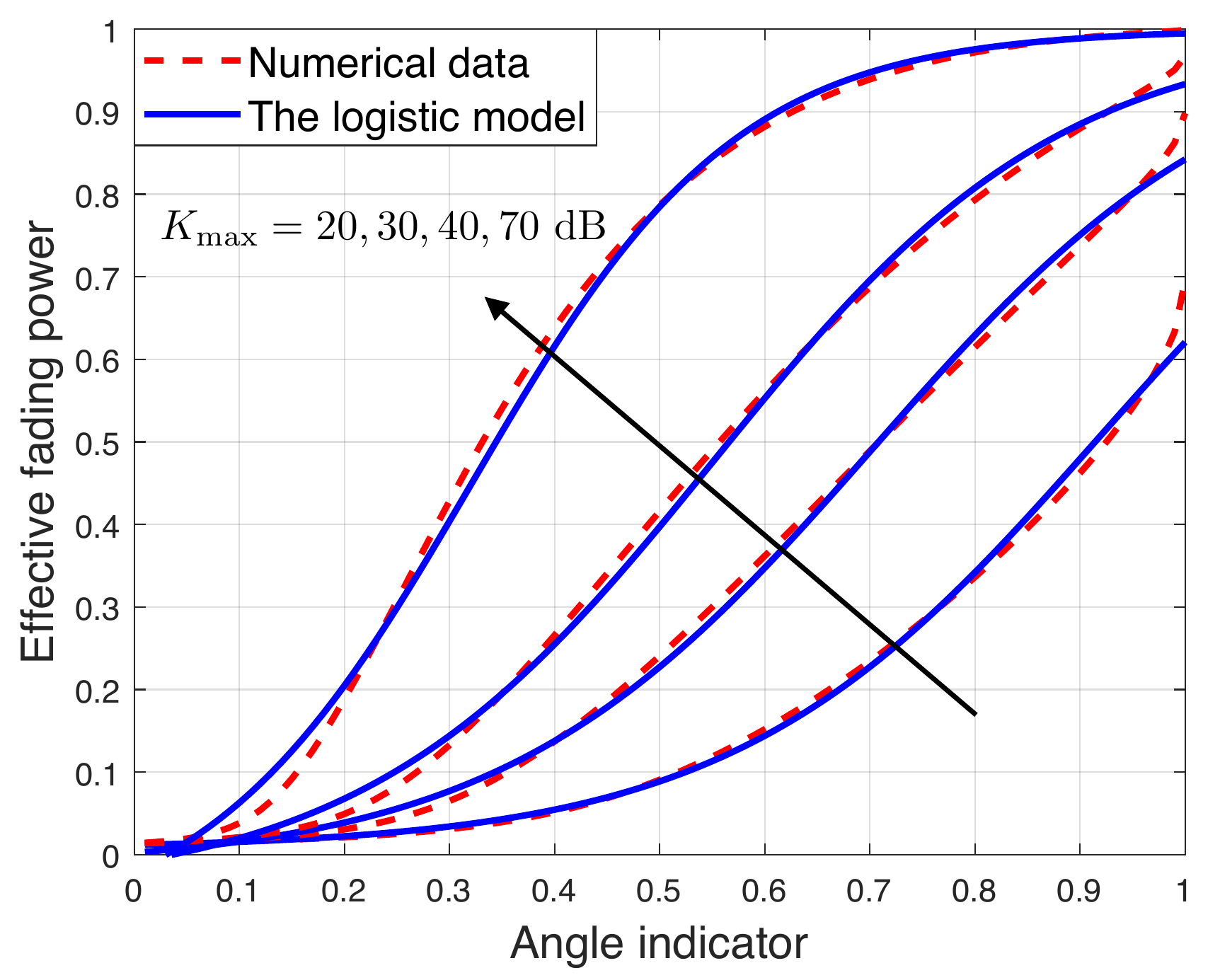}}
\hspace{10mm}
\subfigure[Maximum tolerable outage probability $=0.1$.]{\label{Fig:Regression4}
\includegraphics[height=5.3cm]{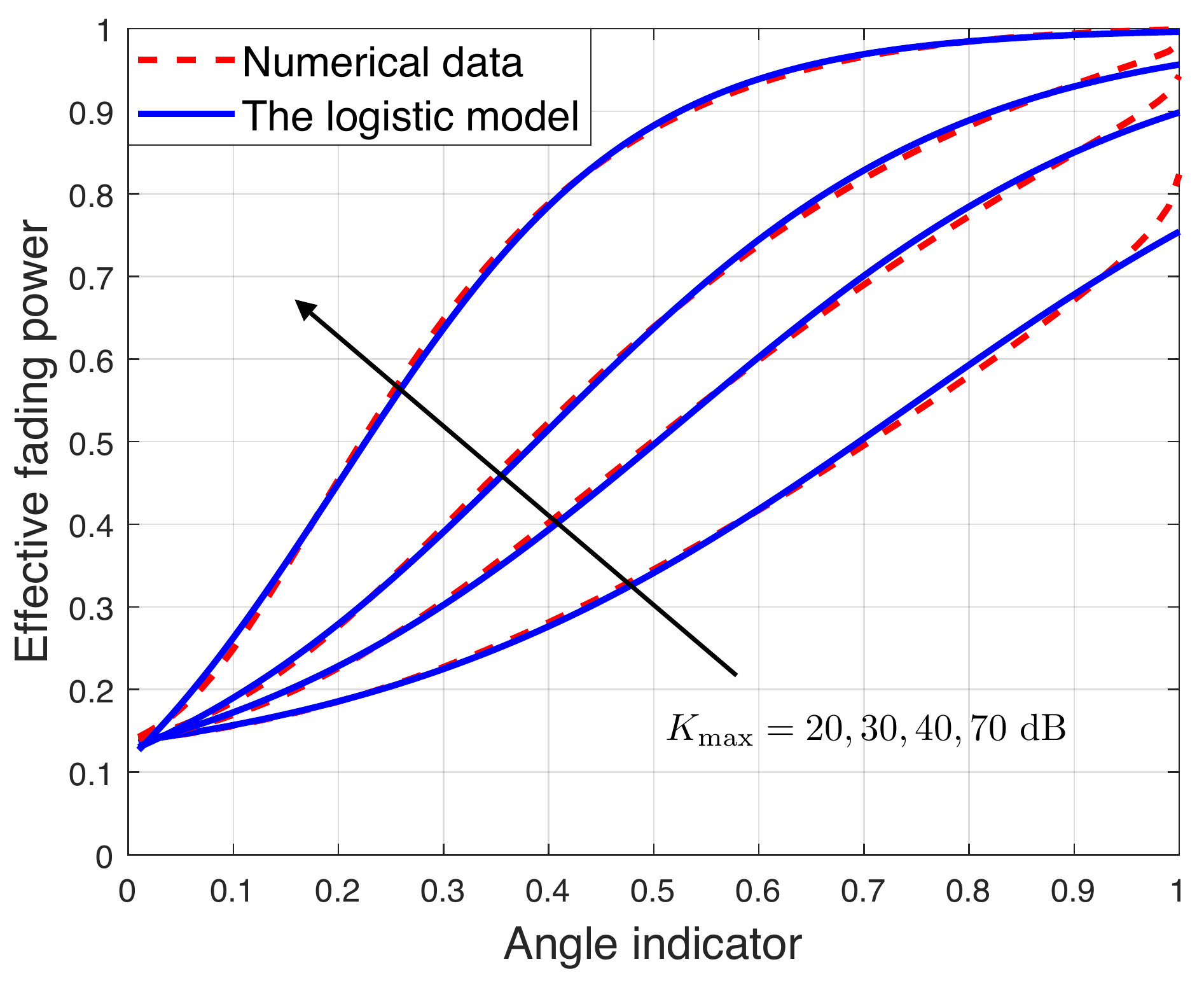}}
\caption{Comparisons between numerical data and the logistic model with $K_{\min}=0$ dB.}
\label{Fig:Regression}
\end{figure} 

To overcome this difficulty, in this work, we apply the \emph{data regression} method to \emph{directly} approximate the effective-fading-power function w.r.t. the 3D UAV trajectory, which comprises the following two key procedures.
\begin{itemize}
\item[1.] {\bf{Model selection:}} We first generate the numeral data for $f$ according to 
 \eqref{Eq:Distance}--\eqref{Eq:angle}, \eqref{Eq:OutProb} and \eqref{Eq:cdf}, and plot the curves of $f$ versus (vs.) $v$ under different maximum Rician factors  and maximum tolerable outage probabilities as shown in Fig.~\ref{Fig:Regression} (dash lines), where $v$ is called \emph{angle indicator} defined as $v\overset{\triangle}{=}\sin(\theta)=\frac{z}{\sqrt{|| {\mathbf{q}}-{\mathbf{w}}||^2+z^2}}$. 
Several important observations are listed as follows. 
\begin{itemize}
\item[(i)] $f_{\min}\le f\le 1$, where $0\le f_{\min}\le 1$.
\item[(ii)] $f$ is monotonically non-decreasing with $v$.
\item[(iii)] For large $K_{\max}$, as $v$ increases, the derivative of $f$ would increase to a maximum value and then decrease.
\end{itemize}
The above observations indicate that the effective-fading-power function should have an `S' shape w.r.t. $v$ and within the range of $[f_{\min}, 1]$. This suggests to approximate the effective-fading-power function by the following logistic model:
\begin{align}
f\approx\tilde{f}&=\tilde{\phi}(\mathbf{q}, z)\nn\\
&\overset{\triangle}{=}C_1+\frac{C_2}{1+e^{-(B_1+B_2 v)}} \nn \\
&=C_1+\frac{C_2}{1+\exp\l(-\l(B_1+B_2 \dfrac{z}{\sqrt{|| {\mathbf{q}}-{\mathbf{w}}||^2+z^2}}\r)\r)},\label{Eq:GRegres}
\end{align} 
where the coefficients $B_1<0$ reflects the positive logistic mid-point, $B_2>0$ is the logistic growth rate, $C_1>0$ and $C_2>0$  satisfy $C_1+C_2=1$.
\item[2.] {\bf{Model evaluation:}} Next, to evaluate the logistic model, we choose the model parameters based on the criterion of minimum mean square error and fit the model to the numerical data as shown in Fig.~\ref{Fig:Regression} (solid lines). It can be observed that the proposed logistic model in \eqref{Eq:GRegres} matches the numerical data in most cases. 
\end{itemize}

Compared with \eqref{Eq:GAppro}, the newly approximated effective-fading-power function in \eqref{Eq:GRegres} is much more tractable and thus can  facilitate characterizing its relationship with the 3D UAV trajectory ($\tilde{f}$ is also called effective fading power  in the sequel for brevity without causing confusion). In particular, if the horizontal distance is much larger than the UAV altitude, we have $v\approx0$ and the smallest effective fading power is $\tilde{f}_{\min}=C_1+\frac{C_2}{1+e^{-B_1}}$. On the other hand, if the UAV is right above the SN, i.e., $|| {\mathbf{q}}-{\mathbf{w}}||=0$, we have $v=1$ and $\tilde{f}$ achieves its maximum value $\tilde{f}_{\max}=C_1+\frac{C_2}{1+e^{-(B_1+B_2)}}$. In other cases, the effective fading power increases with the angle indicator $v$ following the `S' shape. This shape indicates that in the small elevation-angle regime, slightly enlarging the elevation angle can significantly enhance the effective fading power, but the improvement  diminishes after the angle indictor exceeds a certain threshold.

The specific logistic model is essentially determined by the maximum tolerable outage probability and Rician factor coefficients. Their effects on the effective fading power can be observed from Fig.~\ref{Fig:Regression}, which are summarized as follows. 
\begin{itemize}
\item[1.] \textbf{Effects of maximum tolerable outage probability:} 
First, given the elevation angle and Rician factor coefficients $\{K_{\min},K_{\max}\}$, a larger maximum tolerable outage probability results in a higher effective fading power. This implies that the SN can transmit data at a higher rate when the outage probability requirement relaxes. Second, comparing Figs.~\ref{Fig:Regression3} and \ref{Fig:Regression4}, we can observe that for the case with a more tolerable  outage requirement, it has a larger minimum effective fading power at $v=0$ and thus the improvement on the effective fading power by increasing the elevation angle is more limited.
\item[2.] \textbf{Effects of maximum Rician factor:} 
We can observe from both Figs.~\ref{Fig:Regression3} and \ref{Fig:Regression4} that for the case with a larger maximum Rician factor $K_{\max}$, the effective fading power grows faster in the small elevation-angle regime and  saturates to its maximum value earlier. This infers that if $K_{\max}\to \infty$, the Rician fading can be approximated as the LoS channel and  we have $\tilde{f}\approx1$ for almost all elevation angles. On the other hand, if $K_{\max}= K_{\min}\to0$, the Rician fading reduces to Rayleigh fading and $\tilde{f}=C_1+\frac{C_2}{1+e^{-B_1}}$ for all elevation angles. Therefore, in these two extreme cases, enlarging the elevation angle cannot bring significant improvement on the effective fading power.
\end{itemize}

\subsection{Problem Reformulation} 
Based on the logistic modelling for the effective-fading-power function, the achievable rate given the prescribed outage probability requirement, $R_n[m]$ in \eqref{Eq:Rate}, can be approximated as
\begin{equation}
R_n[m]\approx\tilde{R}_n[m]\overset{\triangle}{=}\log_2\l(1+\l(C_1+\frac{C_2}{1+e^{-\l(B_1+B_2 v_n[m]\r)}}\r)\l(\frac{ \gamma_n}{ {\color{black} (|| \mathbf{q}[m]-\mathbf{w}_n||^2+z[m]^2)^{\alpha/2}}}\r)\r), \label{Eq:ApprRate}
\end{equation}
and thus Problem P1 is readily reformulated as:
\begin{subequations}
%\tag{$\textbf{P2}$} 
\begin{align}({\bf P2})
\max_{\l\{\substack{\mathbf{q}[m], z[m] \\ a_n[m], v_n[m]}\r\}, \eta} ~&\eta \nn \\  
\text{s.t.}~~&\frac{1}{M}\sum_{m=1}^{M} a_n[m] \tilde{R}_n[m]\ge \eta, &&\forall n, \label{Eq:P2RateCons}\\
&v_n[m]=\frac{z[m]}{\sqrt{|| {\mathbf{q}}[m]-{\mathbf{w}}_n||^2+z[m]^2}}, && \forall n, m,\label{Eq:P2vcons}\\
&\eqref{Eq:P1ConsStar}-\eqref{Eq:P1InterCons} \nn,
\end{align}
\end{subequations}
where $\tilde{R}_n[m]$ is expressed in \eqref{Eq:ApprRate}.
Problem P2 is still difficult to solve due to its non-convexity that arises from the coupling among $v_n[m]$, $\mathbf{q}_n$ and $z[m]$ in the rate constraint \eqref{Eq:P2RateCons}, the non-affine equality constraint \eqref{Eq:P2vcons}, and the integer UAV scheduling constraint \eqref{Eq:P1InterCons}. To tackle these difficulties, an efficient algorithm is proposed in the next section for attaining a suboptimal solution to Problem P2.

\section{Proposed Algorithm for Problem P2}\label{Sec:PropAlgo}
To solve Problem P2, we first relax the integer constraint for the UAV scheduling in \eqref{Eq:P1InterCons}, leading to the following optimization problem:
\begin{subequations}
%\tag{$\textbf{P2}$} 
\begin{align}({\bf P3})
\max_{\l\{\substack{\mathbf{q}[m], z[m] \\ a_n[m], v_n[m]}\r\}, \eta} ~~~&\eta \nn\\  
\text{s.t.}\quad
&0\le a_n[m]\le1, \qquad\qquad\forall n,m,\label{Eq:P4InterRelax}\\
&\eqref{Eq:P1ConsStar}-\eqref{Eq:ScheCons}, \eqref{Eq:P2RateCons}, \eqref{Eq:P2vcons}.\nn
\end{align}
\end{subequations}
Observe that this relaxed problem is still non-convex. Among others, one of the major challenges for solving it is the non-affine equality constraint \eqref{Eq:P2vcons}. To address it, one important property of Problem P3 is presented as below, which will facilitate the subsequent optimizations.
\begin{lemma}\label{Lem:OptRelax}
\emph{The solution to Problem P3 can be obtained by solving Problem P4 formulated below that relaxes the equality constraint \eqref{Eq:P2vcons} as the inequality constraint:
\begin{subequations}
\begin{align}({\bf P4})
\max_{\l\{\substack{\mathbf{q}[m], z[m] \\ a_n[m], v_n[m]}\r\}, \eta} ~~~&\eta \nn \\  
\text{s.t.}~~&v_n[m]\le\frac{z[m]}{\sqrt{|| {\mathbf{q}}[m]-{\mathbf{w}}_n||^2+z[m]^2}}, && \forall n, m,\label{Eq:P4vcons}\\
&\eqref{Eq:P1ConsStar}-\eqref{Eq:ScheCons}, \eqref{Eq:P2RateCons}, \eqref{Eq:P4InterRelax}.\nn
\end{align}
\end{subequations}
}
\end{lemma}
\begin{proof}
See Appendix~\ref{App:OptRelax}.
\end{proof}

The equivalent Problem P4 remains non-convex due to the existence of coupling variables in the constraints. To address this challenge, we propose to derive a suboptimal  solution to Problem P4 by applying the BCD and SCA techniques. Specifically, given the 3D UAV  trajectory $\{\mathbf{q}[m], z[m]\}$, the UAV scheduling is optimized by solving a linear programming (LP). For any feasible UAV scheduling and vertical trajectory, we optimize  the UAV horizontal trajectory by using the SCA technique. The approach is also applied to optimize the UAV vertical trajectory given any feasible UAV scheduling and horizontal trajectory.  These subproblems are solved in the following subsections. Last, we summarize the overall algorithm and its convergence property.

\subsection{UAV Communication Scheduling Optimization}
Given any feasible 3D UAV trajectory $\{\mathbf{q}[m], z[m]\}$, Problem P4 reduces to:
\begin{subequations}
%\tag{$\textbf{P2}$} 
\begin{align}({\bf P5})
\max_{\{a_n[m]\},  \eta } ~~~&\eta \nn\\
\text{s.t.}\quad
 &\eqref{Eq:ScheCons}, \eqref{Eq:P2RateCons}, \eqref{Eq:P4InterRelax}.\nn
\end{align}
\end{subequations}
Problem P5 is a standard LP which can be efficiently solved by existing solvers, e.g., CVX. Moreover, it can be proved by contradiction that in the optimal solution to Problem P5, the constraints on the UAV scheduling in \eqref{Eq:ScheCons} for all time slots are active, i.e., $\sum_{n=1}^{N} a_n^*[m]= 1, \forall m.$

\subsection{UAV Horizontal Trajectory Optimization}\label{Sec:HorizOpt}
Given any feasible UAV scheduling $\{a_n[m]\}$ and its vertical trajectory $\{z[m]\}$, Problem P4 can be rewritten as the following optimization problem for the UAV horizontal trajectory:
\begin{subequations}
%\tag{$\textbf{P2}$} 
\begin{align}({\bf P6})
\max_{\l\{\mathbf{q}[m], v_n[m]\r\}, \eta} ~~~&\eta \nn \\  
\text{s.t.}~~
& {\mathbf q}[1]={\mathbf{q}}_0, \quad {\mathbf q}[M+1]={\mathbf{q}}_F, \label{Eq:P6StarEnd} \\
& \eqref{Eq:P1ConsStar},\eqref{Eq:P2RateCons}, \eqref{Eq:P4vcons}.\nn
\end{align}
\end{subequations}
First, observe that in the  rate constraint \eqref{Eq:P2RateCons}, $\tilde{R}_n[m]$ given in \eqref{Eq:ApprRate}  is neither concave nor convex w.r.t. the optimization variables $v_n[m]$ and $\mathbf{q}[m]$. To tackle this difficulty, we first introduce an important lemma as below.
\begin{lemma}\label{Lem:ConvXY}\emph{
Given $\gamma, C_1, C_2\ge0$, the function {\color{black} $\psi(x,y)\overset{\triangle}{=}\log_2\l(1+\l(C_1+\frac{C_2}{x}\r)\frac{\gamma}{y^{\alpha/2}}\r)$} is convex w.r.t. $x>0$ and $y>0$.}
\end{lemma}
\begin{proof}
See Appendix~\ref{App:ConvXY}.
\end{proof}

Using Lemma~\ref{Lem:ConvXY}, we can easily prove that $\tilde{R}_n[m]$ in \eqref{Eq:ApprRate} is a convex function w.r.t. $(1+e^{-\l(B_1+B_2 v_n[m]\r)})$ and $(|| \mathbf{q}[m]-\mathbf{w}_n||^2+z[m]^2)$. Although the constraint  \eqref{Eq:P2RateCons} is still non-convex, we can leverage the SCA technique to derive its convex approximation. To be specific, using the fact that  the first-order Taylor approximation of a convex function is a global under-estimator, $\tilde{R}_n[m]$ can be lower-bounded as follows.
\begin{lemma}\label{Lem:RateBound}\emph{
For any local UAV horizontal trajectory, $\{\hat{\mathbf{q}}_n[m]\}$, we have
\begin{equation}
\tilde{R}_n[m]\ge \tilde{R}_n^{\rm{lb}}[m]\overset{\triangle}{=}\hat{R}_n[m]-\hat{\Phi}_n[m] (e^{-s_n[m]}-e^{-\hat{s}_n[m]})-\hat{\Psi}_n[m](|| \mathbf{q}[m]-\mathbf{w}_n||^2-|| \hat{\mathbf{q}}[m]-\mathbf{w}_n||^2), \forall n,m,\nn
\end{equation}
where the equality holds at the point $\mathbf{q}[m]=\hat{\mathbf{q}}[m]$. The coefficients $\hat{R}_n[m]$, $\hat{\Psi}_n[m]$, $\hat{\Phi}_n[m]$, and $\hat{s}_n[m]$ are defined in Appendix~\ref{App:RateBound}, and $s_n[m]$ is defined by
\begin{equation}
s_n[m]\overset{\triangle}{=}B_1+B_2 v_n[m].\label{Eq:s}
\end{equation}}
\end{lemma}
\begin{proof}
See Appendix~\ref{App:RateBound}.
\end{proof}

Modifying Problem P6 by replacing $\tilde{R}_n[m]$ given in \eqref{Eq:ApprRate}  with its lower bound in Lemma~\ref{Lem:RateBound}, $\tilde{R}_n^{\rm{lb}}[m]$, and combing \eqref{Eq:P4vcons} with \eqref{Eq:s} yields the following approximate problem:
\begin{subequations}
%\tag{$\textbf{P2}$} 
\begin{align}({\bf P7})
\max_{\l\{\mathbf{q}[m], s_n[m]\r\}, \eta} ~&\eta\nn \\  
\text{s.t.}\quad&\frac{1}{M}\sum_{m=1}^{M} a_n[m]  \tilde{R}_n^{\rm{lb}}[m]
%\l(\hat{R}_n[m]-\hat{\Phi}_n[m] (e^{-s_n[m]}-e^{-\hat{s}_n[m]})\right.\nn\\ 
%&\qquad\quad \left.-\hat{\Psi}_n[m](|| \mathbf{q}[m]-\mathbf{w}_n||^2-|| \hat{\mathbf{q}}[m]-\mathbf{w}_n||^2)\r)
\ge \eta, &&\forall n,\label{Eq:P7Rbound}\\
& s_n[m]\le B_1+B_2 \frac{z[m]}{\sqrt{|| {\mathbf{q}}[m]-{\mathbf{w}}_n||^2+z[m]^2}}, &&\forall n, m\label{Eq:P7Equality}\\
& \eqref{Eq:P1ConsStar}, \eqref{Eq:P6StarEnd}.\nn
\end{align}
\end{subequations}
The remaining difficulty for solving Problem P7 is the constraint \eqref{Eq:P7Equality}, for which $v_n[m]=\frac{z[m]}{\sqrt{|| {\mathbf{q}}[m]-{\mathbf{w}}_n||^2+z[m]^2}}$ is not concave w.r.t. ${\mathbf{q}}[m]$. To address it, one key observation is that $v_n[m]$ is convex  w.r.t. $(|| {\mathbf{q}}[m]-{\mathbf{w}}_n||^2+z[m]^2)$. This useful property allows us to lower-bound $v_n[m]$ by using the SCA technique, which is given as follows.

\begin{lemma}\label{Lem:vSCA}\emph{
For any local UAV horizontal trajectory, $\{\hat{\mathbf{q}}_n[m]\}$, we have 
\begin{equation}
v_n[m]\ge v_n^{\rm{lb}}[m]\overset{\triangle}{=}\hat{v}_n[m]-\hat{\Lambda}_n[m](|| {\mathbf{q}}[m]-{\mathbf{w}}_n||^2-|| {\hat{\mathbf{q}}}[m]-{\mathbf{w}}_n||^2), \quad\forall n,m, \label{Eq:vBound}
\end{equation}
where the equality holds at the point $\mathbf{q}[m]=\hat{\mathbf{q}}[m]$, and the coefficients $\hat{v}_n[m]$ and $\hat{\Lambda}_n[m]$ are defined in Appendix~\ref{App:vSCA}.
}
\end{lemma} 
\begin{proof}
See Appendix~\ref{App:vSCA}.
\end{proof}

Consequently, Problem P7 can be transformed to the following approximate problem by substituting \eqref{Eq:vBound} into \eqref{Eq:P7Equality}:
\begin{subequations}
%\tag{$\textbf{P2}$} 
\begin{align}({\bf P8})
\max_{\l\{\mathbf{q}[m], s_n[m]\r\}, \eta} ~&\eta \nn\\  
\text{s.t.}\quad
& s_n[m]\le B_1\!+\!B_2 v_n^{\rm{lb}}[m]
%\l(\hat{v}_n[m]\!-\!\hat{\Lambda}_n[m](|| {\mathbf{q}}[m]\!\!-\!\!{\mathbf{w}}_n||^2\!-\!|| {\hat{\mathbf{q}}}[m]\!-\!{\mathbf{w}}_n||^2)\r)
, \!\!&&\forall n, m,\nn\\
&\eqref{Eq:P1ConsStar}, \eqref{Eq:P6StarEnd}, \eqref{Eq:P7Rbound}.\nn
\end{align}
\end{subequations}
Problem P8 is now a convex optimization problem, which can be efficiently solved by using existing solvers, e.g., CVX. It is worthwhile to note that, by approximating the concave constraints with their convex lower bounds, the feasible set of  Problem P8 is always a subset of Problem P6. Therefore, solving Problem P8 gives the lower bound of the objective value in Problem P6.

\subsection{UAV Vertical Trajectory Optimization}
Given any feasible UAV scheduling $\{a_n[m]\}$ and its horizontal trajectory $\{\mathbf{q}[m]\}$, Problem P4 can be rewritten as  the following optimization problem for the UAV vertical trajectory:
\begin{subequations}
%\tag{$\textbf{P2}$} 
\begin{align}({\bf P9})
\max_{\l\{z[m],  v_n[m]\r\}, \eta} ~~~&\eta  \nn\\  
\text{s.t.}\quad& z[1]=z_0, ~~~z[M+1]=z_F,\label{Eq:P9z}\\
&\eqref{Eq:P1VerSpeed}, \eqref{Eq:zmin}, \eqref{Eq:P2RateCons}, \eqref{Eq:P4vcons}.\nn
\end{align}
\end{subequations}
Observe that Problem P9 has a similar form with Problem P6. Therefore, following a similar procedure as for solving Problem P6 (i.e., applying the SCA technique for the constraint \eqref{Eq:P2RateCons}), Problem P9 can be transformed to the following approximate problem.
\begin{subequations}
\begin{align}({\bf P10})
\max_{\l\{z[m], s_n[m]\r\}, \eta} ~~&\eta  \nn\\ 
\text{s.t.}~~&\frac{1}{M}\sum_{m=1}^{M} a_n[m]  \l(\check{R}_n[m]-\check{\Phi}_n[m] (e^{-s_n[m]}-e^{-\check{s}_n[m]})\right.\nn\\ 
&\qquad\quad \left.-\check{\Psi}_n[m](z[m]^2- \check{z}[m]^2)\r)\ge \eta, &&\forall n, \label{Eq:P10RateCons}\\
&s_n[m]\le B_1+B_2 \frac{z[m]}{\sqrt{(|| {{\mathbf{q}}}[m]-{\mathbf{w}}_n||^2+z[m]^2)}}, && \forall n, m, \\
&\eqref{Eq:P1VerSpeed}, \eqref{Eq:zmin}, \eqref{Eq:P9z},\nn
\end{align}
\end{subequations}
where $\check{R}_n[m]$, $\check{\Phi}_n[m]$, $\check{s}_n[m]$ and $\check{z}_n[m]$ are the coefficients determined by the local vertical trajectory $\{\hat{z}[m]\}$ and can be derived using the similar method as in Appendix~\ref{App:RateBound} and thus omitted for brevity.
As proved in Appendix~\ref{App:ConvP10}, Problem P10 is a convex optimization problem and thus it can be readily solved by using existing methods, e.g., the interior-point method.\footnote{In practice, due to the lack of support for the function of $v_n[m]$ in CVX, we can also approximate  $v_n[m]$ by its upper bound using the first-order Taylor expansion for simplicity.}

\subsection{Overall Algorithm, Complexity, and Convergence}
\begin{algorithm}[t]
  \caption{Proposed algorithm for Problem P2.}
  \label{Alg:SolveP2}
   \begin{algorithmic}[1]
   \STATE Initialize $\{\mathbf{q}[n], z[n]\}$. Let $i=0$.
\REPEAT
\STATE Solve Problem P5 for given $\{\mathbf{q}^{i}[m], z^{i}[m]\}$, and denote the optimal solution as $\{\alpha^{i+1}_n[m]\}$.
\STATE  Solve Problem P8 for given $\!\{a_n^{i+1}[m], z^{i}[m]\}\!$, and denote the optimal solution as $\!\{\mathbf{q}^{i+1}[m]\}$.
\STATE  Solve Problem P10 for given $\!\{\!a_n^{i+1}\![m],\! \mathbf{q}^{i+1}[m]\!\}\!$, and denote the optimal solution as $\!\{\!z^{i+1}\![m]\!\}\!$. 
\STATE Updata $i=i+1$.
   \UNTIL{Converge to a prescribed accuracy.}
  \end{algorithmic}
  \end{algorithm}
  
 Using the results obtained in the previous three subsections, the overall algorithm for computing the suboptimal  solution to Problem P2 is summarized in Algorithm~\ref{Alg:SolveP2} with the computation complexity analyzed as follows. In each iteration, the UAV communication scheduling, horizontal trajectory, and vertical trajectory are sequentially optimized using the convex solver based on the  interior-point method, and thus their individual  complexity can be represented by $\mathcal{O}((NM)^{3.5}\log(1/\epsilon))$, $\mathcal{O}((M+NM)^{3.5}\log(1/\epsilon))$, and $\mathcal{O}((M+NM)^{3.5}\log(1/\epsilon))$,  respectively, given the solution accuracy of $\epsilon> 0$ \cite{ben2001lectures}. Then accounting for the BCD iterations with the complexity in the order of $\log(1/\epsilon)$, the total computation complexity of Algorithm~\ref{Alg:SolveP2} is $\mathcal{O}((M+NM)^{3.5}\log^2(1/\epsilon))$.

 Next, we address the convergence of Algorithm~\ref{Alg:SolveP2}. Let $\eta(\{{a_n^{i}[m]}\}, \{\mathbf{q}^{i}[m]\}, \{z^{i}[m]\})$ denote the  objective value of Problem P2 in the $i$-th iteration. Since Problem P5 is optimally solved, we have 
 \begin{equation}
 \eta(\{{a_n^{i}[m]}\}, \{\mathbf{q}^{i}[m]\}, \{z^{i}[m]\})\le \eta_{\alpha}(\{{a_n^{i+1}[m]}\}, \{\mathbf{q}^{i}[m]\}, \{z^{i}[m]\})\label{Eq:CommIncre}
 \end{equation} where $\eta_{\alpha}(\{{a_n^{i+1}[m]}\}, \{\mathbf{q}^{i}[m]\}, \{z^{i}[m]\})$ denotes the computed objective value of Problem P5. For the optimization of UAV horizontal trajectory, we have 
 \begin{align} \label{Eq:HorizIncre}
 &\eta(\{{a_n^{i+1}[m]}\}, \{\mathbf{q}^{i}[m]\}, \{z^{i}[m]\})\overset{(a)} {=}\eta^{\text{lb}}_{\mathbf{q}}(\{{a_n^{i+1}[m]}\}, \{\mathbf{q}^{i}[m]\}, \{z^{i}[m]\})\nn \\
 & \quad \qquad \overset{(b)}{\le}\eta_{\mathbf{q}}^{\text{lb}}(\{{a_n^{i+1}[m]}\}, \{\mathbf{q}^{i+1}[m]\}, \{z^{i}[m]\})
 \overset{(c)}{\le} \eta(\{{a_n^{i+1}[m]}\}, \{\mathbf{q}^{i+1}[m]\}, \{z^{i}[m]\})
 \end{align} 
 where $\eta^{\text{lb}}_{\mathbf{q}}$ denotes the objective value of Problem P8, $(a)$ is due to the tightness of the first-order Taylor expansions at locally points in Problem P8, $(b)$ holds since  P8 is optimally solved, and $(c)$ holds because the optimal objective value of Problem P8 is the lower bound of that of Problem P6 (see Section~\ref{Sec:HorizOpt}). Therefore, solving Problem P8 guarantees that the objective value of Problem P6 is non-decreasing. Using the similar derivation procedure as in  \eqref{Eq:HorizIncre}, we have
\begin{align} \label{Eq:VerticIncre}
 &\eta(\{{a_n^{i+1}[m]}\}, \{\mathbf{q}^{i+1}[m]\}, \{z^{i}[m]\})=\eta^{\text{lb}}_{z}(\{{a_n^{i+1}[m]}\}, \{\mathbf{q}^{i+1}[m]\}, \{z^{i}[m]\})\nn \\
 & \quad \qquad \le\eta_{z}^{\text{lb}}(\{{a_n^{i+1}[m]}\}, \{\mathbf{q}^{i+1}[m]\}, \{z^{i+1}[m]\})
 = \eta(\{{a_n^{i+1}[m]}\}, \{\mathbf{q}^{i+1}[m]\}, \{z^{i+1}[m]\}).
 \end{align} 
Consequently, combing \eqref{Eq:CommIncre}-\eqref{Eq:VerticIncre}, we can obtain that $$\eta(\{{a_n^{i}[m]}\}, \{\mathbf{q}^{i}[m]\}, \{z^{i}[m]\})
\le\eta(\{{a_n^{i+1}[m]}\}, \{\mathbf{q}^{i+1}[m]\}, \{z^{i+1}[m]\}),
$$ which  guarantees that the objective value of Problem P2 is non-decreasing over the iterations and thus Algorithm~\ref{Alg:SolveP2} can converge to a locally optimal solution of Problem P2.

%
% solving Problem P10  for proving the non-decreasing property of optimizing UAV horizontal trajectory as in \eqref{Eq:HorizIncre}  of UAV vertical trajectory Applying the similar derivation procedure for the optimization of UAV vertical trajectory, we can conclude that  solving Problem P8 and P10 guarantees that the objective values of Problem P6 and P9 are non-decreasing over the iterations.
% 
% 
% Therefore, solving Problem P8 guarantees that the objective values of Problem P6  is non-decreasing over the iteration. The similar   is due to and P9 for the UAV horizontal and vertical trajectories, we transform them into their approximate forms as in P8 and P10, whose optimal objective values are the lower bounds of those of the original optimization problems. Mathematically, let $\eta_{\mathbf{q}}(\{{a_n[m]}\}, \{\mathbf{q}[m]\}, \{z[m]\} )$ and $\eta^{\rm{lb}}_{\mathbf{q}}(\{{a_n[m]}\}, \{\mathbf{q}[m]\}, \{z[m]\} )$ respectively denote the objective values of Problem P6 and P8. Thus, we have  Therefore, solving Problem P8 and P10 guarantees that the objective values of Problem P6 and P9 are non-decreasing over the iterations. Combing them with the procedure for solving the convex optimization Problem P5, Algorithm~\ref{Alg:SolveP2} is guaranteed to converge to the locally optimal solution of Problem P2.

Last, it is useful to mention that 1) the initial UAV trajectory can be constructed as the straight flight from the initial location to the final location; and 2) the continuous UAV scheduling obtained from solving Problem P5 can be reconstructed to the binary scheduling using the method in \cite{wu2018joint} without compromising the optimality.

%\begin{remark}\emph{This work can be straightforwardly extended  to account for the limited transmission energy at the SNs by adding the extra constraints on SNs' transmission energy consumption to the original optimization problem.  The corresponding suboptimal solution can be derived by using the similar derivation procedures in the current work except two differences: First, for the new problem, the solution to Problem P4 is only a suboptimal solution to Problem P3; Second, Problem P5 should be modified to account for the constraints on the SNs' energy consumption.}
%\end{remark}

\section{Numerical results}\label{Sec:Nume}
In this section, numerical results are provided to characterize the properties of designed 3D UAV trajectory and evaluate the performance of our proposed algorithm.  We consider a UAV-enabled WSN with SNs randomly and uniformly distributed in a square area of $1000\times1000$ m$^2$. For ease of illustration, the following results are based on one specific realization of SNs' locations for both the cases with single and multiple SNs. Unless otherwise stated, the numerical settings are as follows.  The UAV is assumed to fly from the initial location $(0, 500,100)$ m towards the final location $(1000, 500, 100)$ m within $T=26$ s, with the maximum horizontal and vertical speeds set as $V_{\rm{xy}}=50$ m/s and $V_{\rm{z}}=20$ m/s, respectively. The maximum tolerable outage probability $\epsilon=0.01$ and all the SNs transmit data at the same power of $P_n=0.1$ W.  The channel power gain at the reference distance $d_0=1$ m is $\beta_0=-60$ dB,  {\color{black}the pathloss exponent is $\alpha=2$,} the receive noise power $\sigma^2=-109$ dBm, and the SNR gap $\Gamma=8.2$ dB.  Other parameters are set as $H=100$ m, $\delta=0.2$ s, $K_{\min}=0$ dB, and $K_{\max}=30$ dB.

For comparison, we consider three benchmark schemes, namely, 1) LoS-based (LB) scheme, which designs the UAV scheduling and trajectory assuming the simplified LoS channel as in \cite{wu2018joint}; 2)  Rician-fading lowest altitude (RFLA) scheme that only optimizes the UAV  scheduling and horizontal trajectory proposed in this work without the optimization for the vertical trajectory which is simply set as the lowest altitude, i.e., $100$ m; and 3) Rician-fading fixed suboptimal altitude (RFFSA) scheme, which resembles the RFLA scheme but differs in that it selects the best fixed UAV altitude among several candidate altitudes $\{100, 125, 150, \cdots, 300\}$ m. In addition, our proposed algorithm is named as  Rician-fading based (RFB) scheme. Note that the (average) max-min rates computed from the corresponding algorithms, named as \emph{estimated} max-min rates, are incomparable since in practice they may not be achievable.  For fair comparison, we consider another performance metric, called \emph{achieved} max-min rate, which is computed using the precise outage-aware achievable rate in \eqref{Eq:Rate} and corresponding computed UAV trajectory and scheduling.
%
% This model is inaccurate in the sense that it neglects the angle-dependent Rician fading. Our proposed algorithm is named as  Rician-fading based (RFB) scheme. The (average) max-min rates computed from the corresponding algorithms, named as \emph{estimated} max-min rates, are incomparable since in practice they may not be achievable.  For fair comparison, we consider another performance metric, called \emph{achieved} max-min rate, which is computed using the precise outage-aware achievable rate in \eqref{Eq:Rate} and corresponding computed UAV trajectory and scheduling.

\subsection{Comparisons of Achievable-Rate Functions}
The differences of the optimized UAV trajectories for the LB and RFB schemes essentially arise from their different achievable rates. Compared with the simplified  LoS channel model, the achievable rate for the considered Rician fading model given  in \eqref{Eq:ApprRate} has an extra term, namely the effective fading power, which is determined by the 3D UAV trajectory. The effects of horizontal distance and altitude on the achievable-rate functions are discussed as follows.

\begin{figure}[t!]
\centering
\subfigure[Effects of horizontal distance.]{\label{Fig:HoriTrajec}
\includegraphics[height=5.3cm]{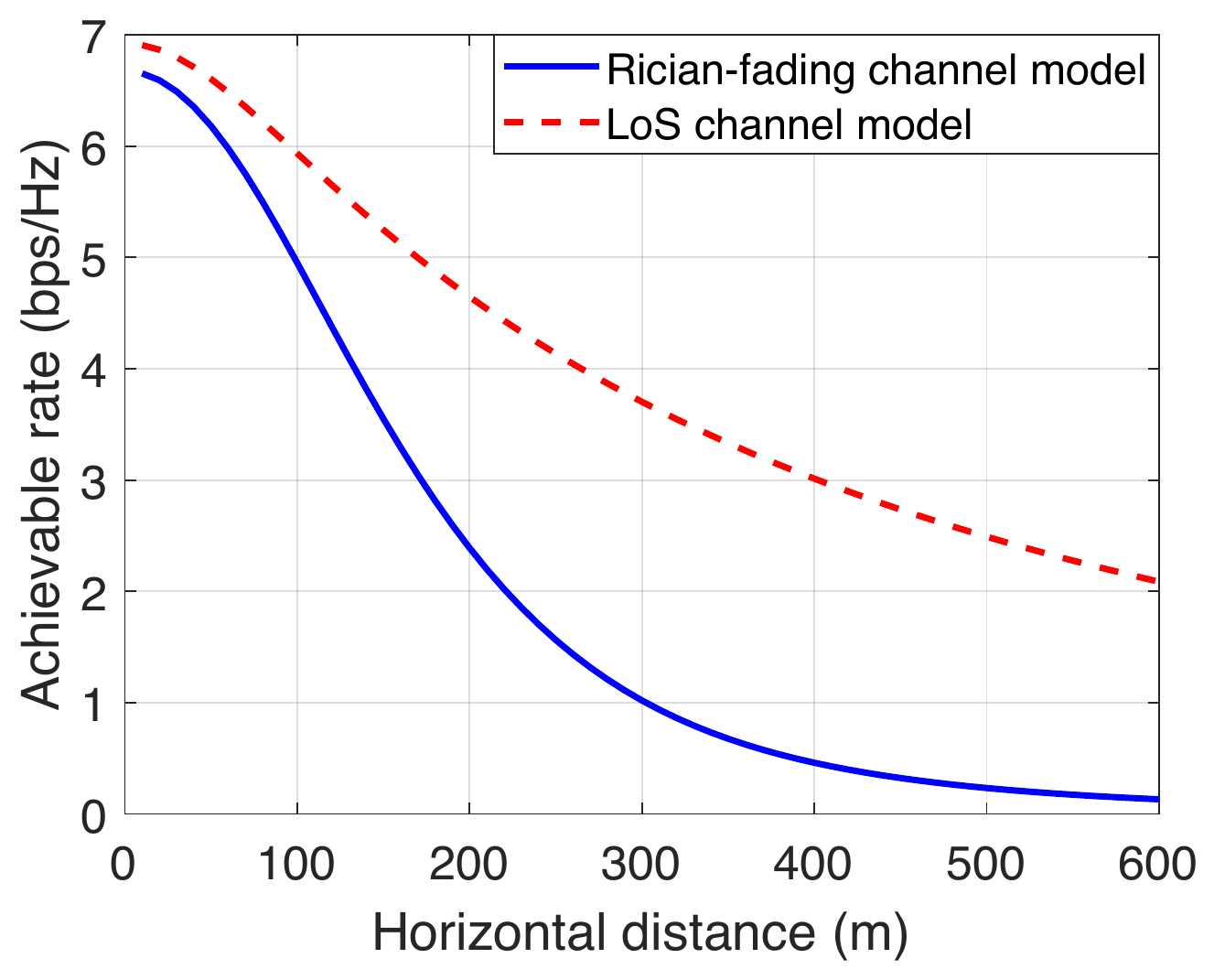}}
\hspace{10mm}
\subfigure[Effects of altitude.]{\label{Fig:VertiTrajec}
\includegraphics[height=5.3cm]{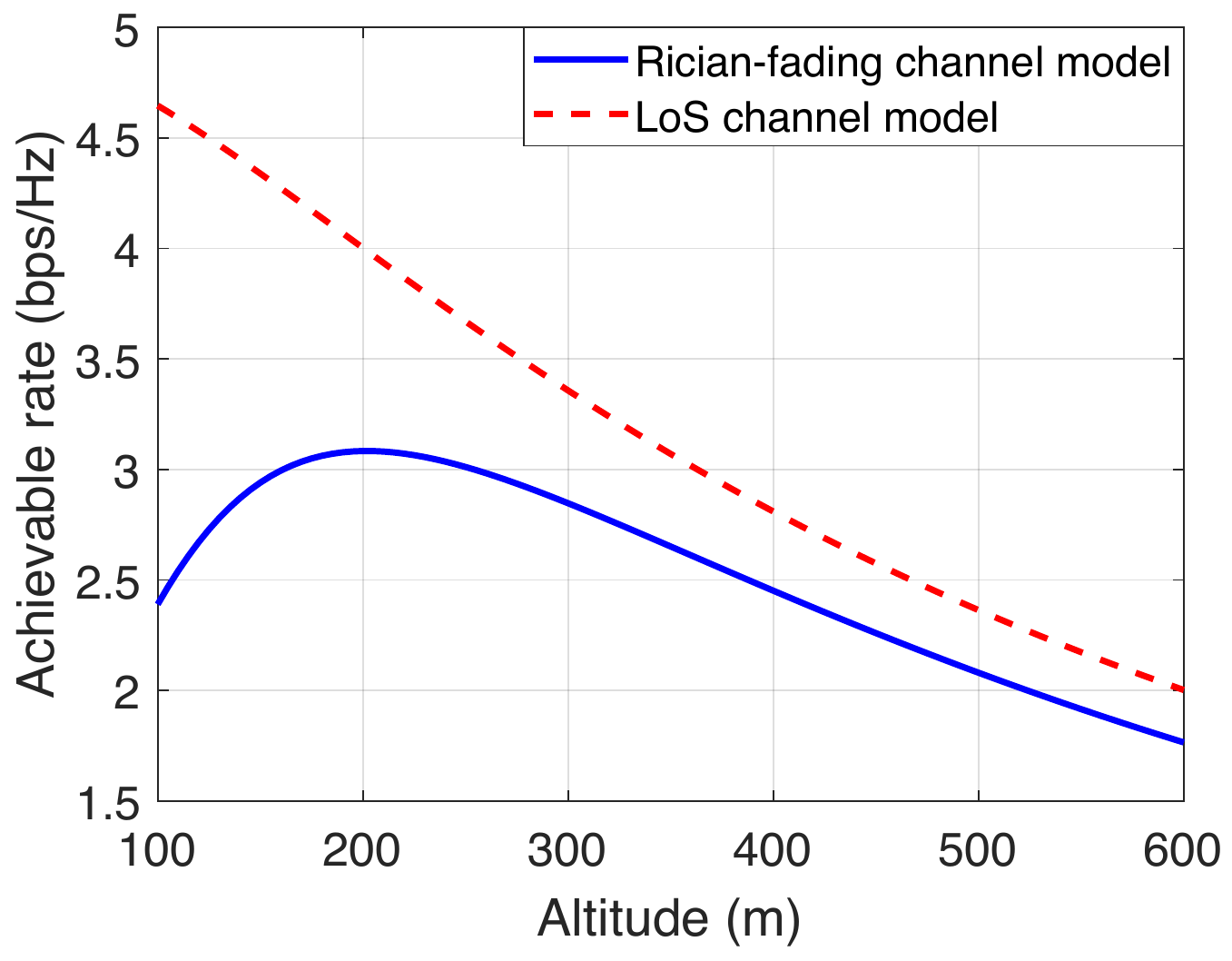}}
\caption{Achievable rates under (a) different horizontal distances and the same altitude ($100$ m); (b) different altitudes and the same horizontal distance ($200$ m) with model parameters $B_1=-4.3221$, $B_2=6.0750$, $C_1=0$, and $C_2=1$.
}
\end{figure}

Fig.~\ref{Fig:HoriTrajec} plots the curves of achievable rate vs. the horizontal distance in different channel models given a fixed UAV altitude. It can be observed that as the horizontal distance increases, the achievable rate for the Rician fading model decreases much \emph{faster} than that of the simplified LoS channel model. This is because for the considered model,  reducing the horizontal distance can not only shorten the UAV-SN distance leading to a smaller pathloss as in the case of LoS channel model, but also enlarge the elevation angle yielding a larger effective fading power.  This difference implies that if given the vertical trajectory, the UAV should fly closer to the scheduled SNs to further reap the \emph{angle gain} by increasing the elevation angle.  Another observation is that the achievable rate under the LoS channel model is always larger than that under the Rician-fading channel model. The reason is that the previous model ensures zero outage probability and the latter, due to the existence of random small-scale  fading, needs to reduce the transmission rate for satisfying the outage probability requirement. 

Fig.~\ref{Fig:VertiTrajec} shows the effects of the UAV altitude on the achievable rates in different channel models. We can observe that, as the altitude increase, the achievable rate under the Rician fading model is firstly increasing and then decreasing, which is significantly different from that under the LoS channel model with a monotonically decreasing rate. The is because except the special case with the UAV right above the SN, raising the UAV to a higher altitude can enlarge the elevation angle leading to a larger effective fading power, but at the same time, result in more pathloss. Consequently, the UAV altitude should be optimized to balance the \emph{pathloss-and-fading} (or \emph{distance-and-angle}) tradeoff in our considered 3D UAV trajectory design.

\subsection{Single SN}
\begin{figure}[t!]
\centering
%\subfigure{\label{Fig:Angle_outage}
\subfigure[3D UAV trajectory.]{\label{FigKTraj}
\includegraphics[height=5.5cm]{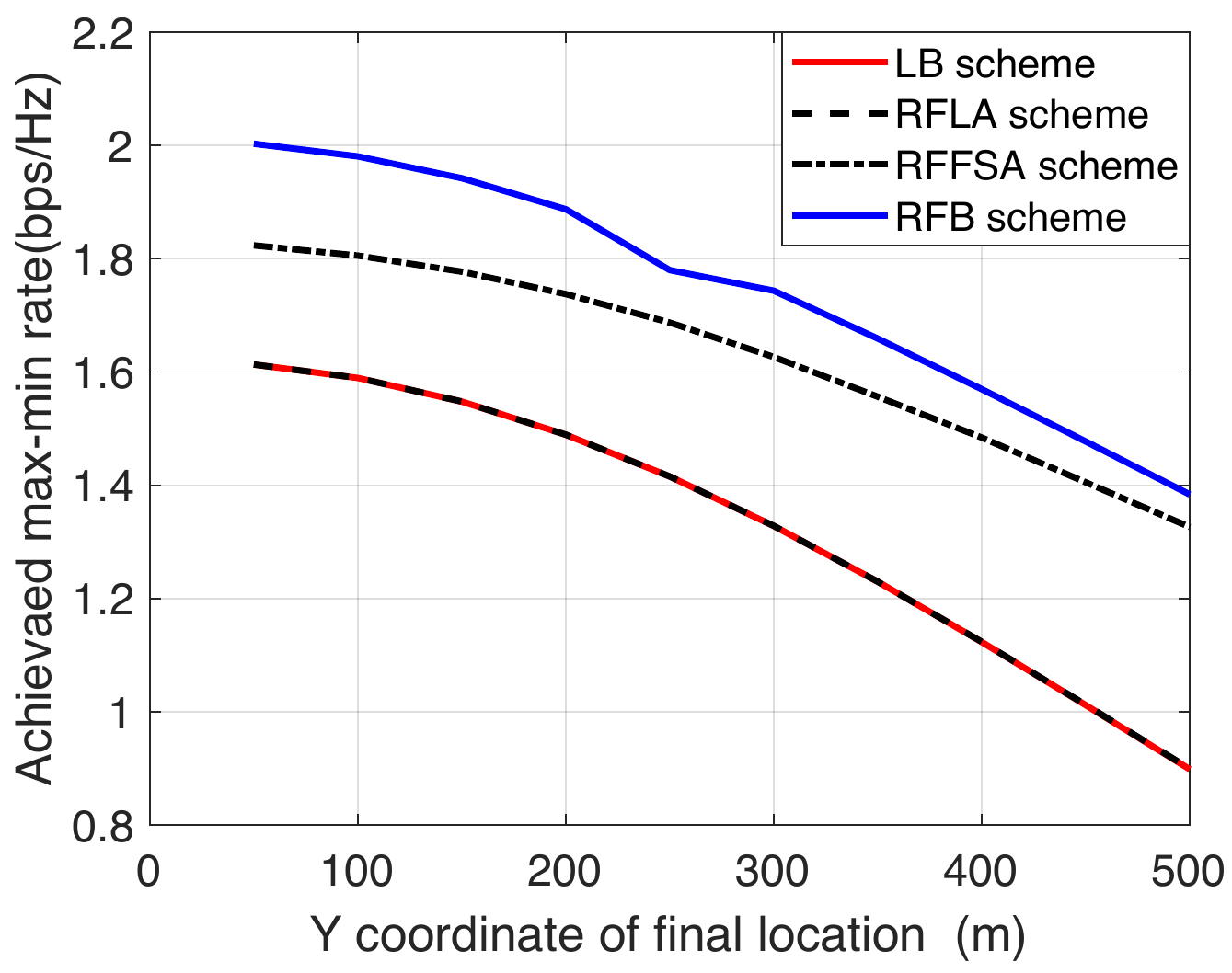}}
\vspace{-0.5mm}
\caption{Effectiveness of proposed algorithm.}\label{Fig:Y}
\end{figure} 
Next, we consider a special case with only one SN located at $(200,0,0)$ and demonstrate the effectiveness of proposed algorithm. Then for easy of illustration, we focus on the comparison of LB and RFB schemes and evaluate the effects of several parameters on the UAV trajectory design and rate  performance, including the time duration, UAV maximum vertical speed, maximum tolerable outage probability, and maximum Rician factor.

 \subsubsection{Effectiveness of proposed algorithm} In Fig.~\ref{Fig:Y}, we evaluate the effectiveness of proposed algorithm by comparing its rate performance with other benchmark schemes under different y coordinates of final location.  First, we can observe that the proposed RFB scheme significantly outperforms the LB scheme in terms of the achieved max-min rate, since it adopts a practical angle-dependent Rician fading model and jointly optimizes the 3D UAV trajectory. Next, as for the RFLA scheme, although it adopts the practical channel model, the UAV vertical trajectory is not jointly optimized with the UAV scheduling and horizontal trajectory, and thereby the scheme cannot fully attain the angle gain as the RFB scheme. Specifically, the RFLA scheme only has marginal rate performance improvement compared with the LB scheme due to similar horizontal trajectory. Third, the RFFSA achieves larger max-min rates than the RFLA scheme as it further optimizes the fixed UAV altitude, but it still suffers considerable performance loss compared with the RFB scheme since it cannot adaptively tune the altitude along the trajectory. These observations show the importance of adopting a practical model and joint 3D trajectory optimization. Last, it is observed that  the rate performance monotonically decreases with the increase of y coordinate of final location, since the UAV has to spend a longer time duration on flying towards the SN and then the destination, and thus has shorter time to collect data.

\subsubsection{Effects of time duration} 
\begin{figure}[t!]
\centering
\subfigure[3D UAV trajectory.]{\label{FigTimeTraj}
\includegraphics[height=5.5cm]{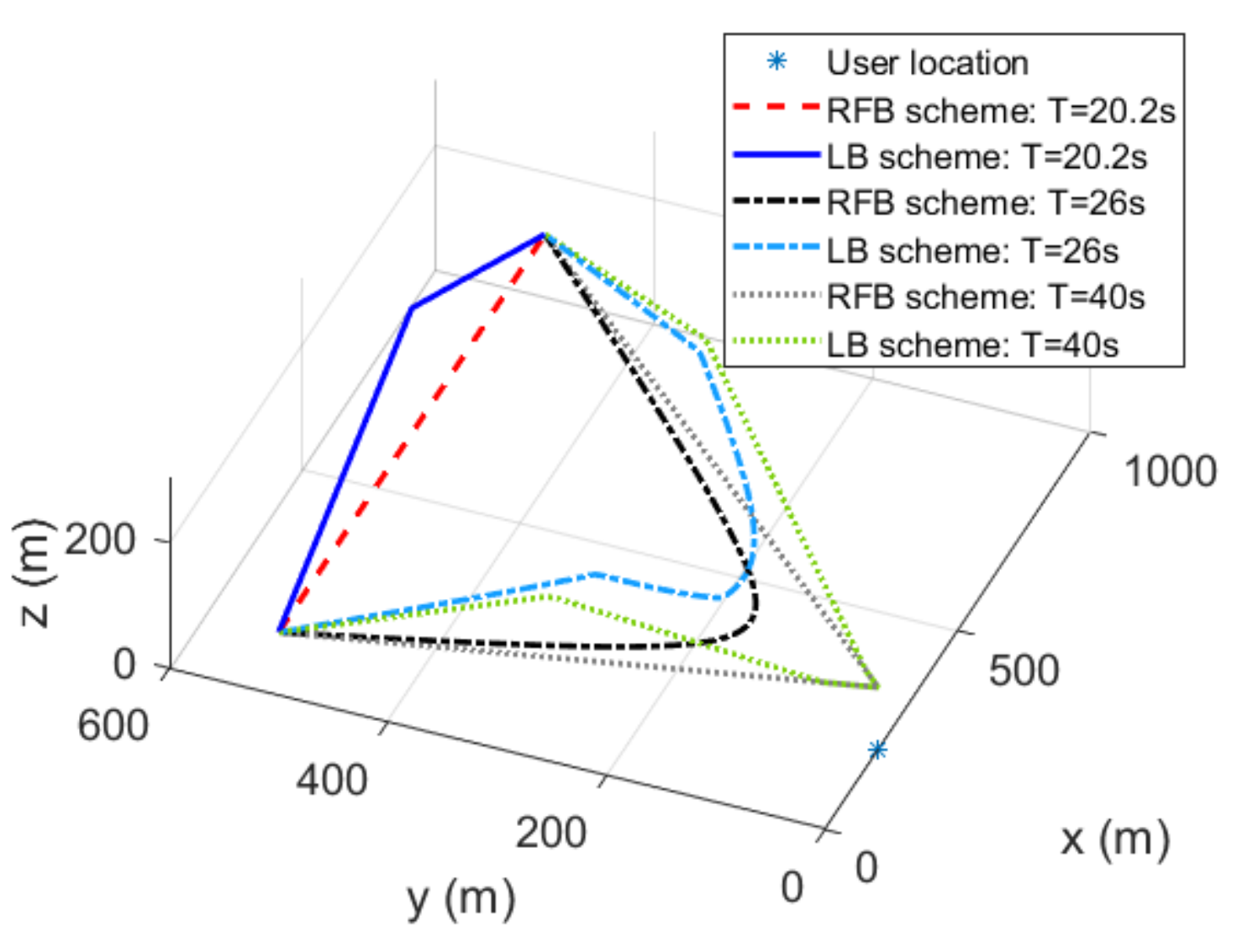}}
%\subfigure[Horizontal UAV trajectory]{\label{FigTimeTrajHoriz}
%\includegraphics[height=6cm]{./FigTimeTrajHoriz.pdf}}
\hspace{10mm}
\subfigure[UAV elevation angle.]{\label{FigTimeTrajAngle}
\includegraphics[height=5.5cm]{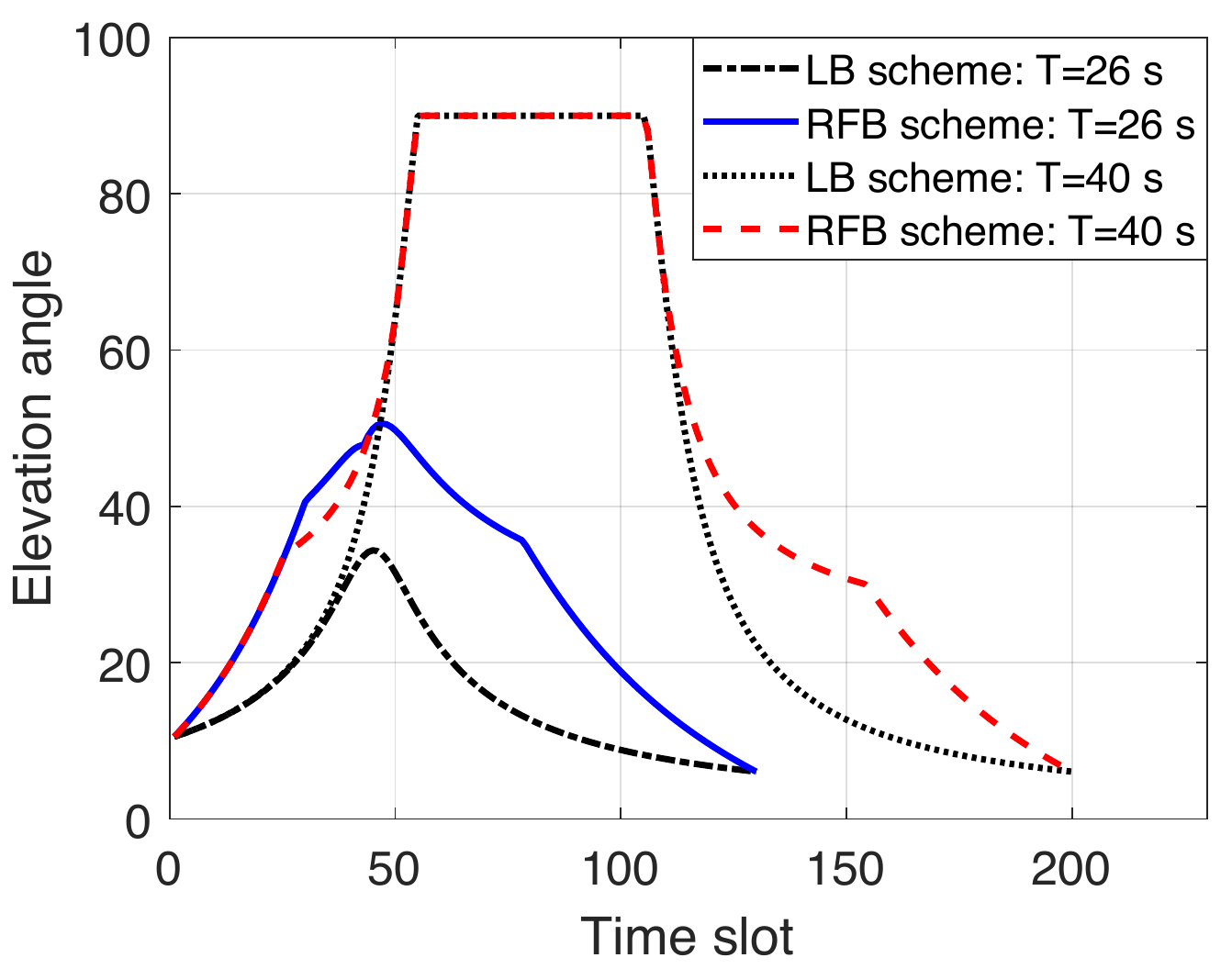}}
\subfigure[Max-min rate  vs. time duration.]{\label{FigTimeRate}
\includegraphics[height=5.5cm]{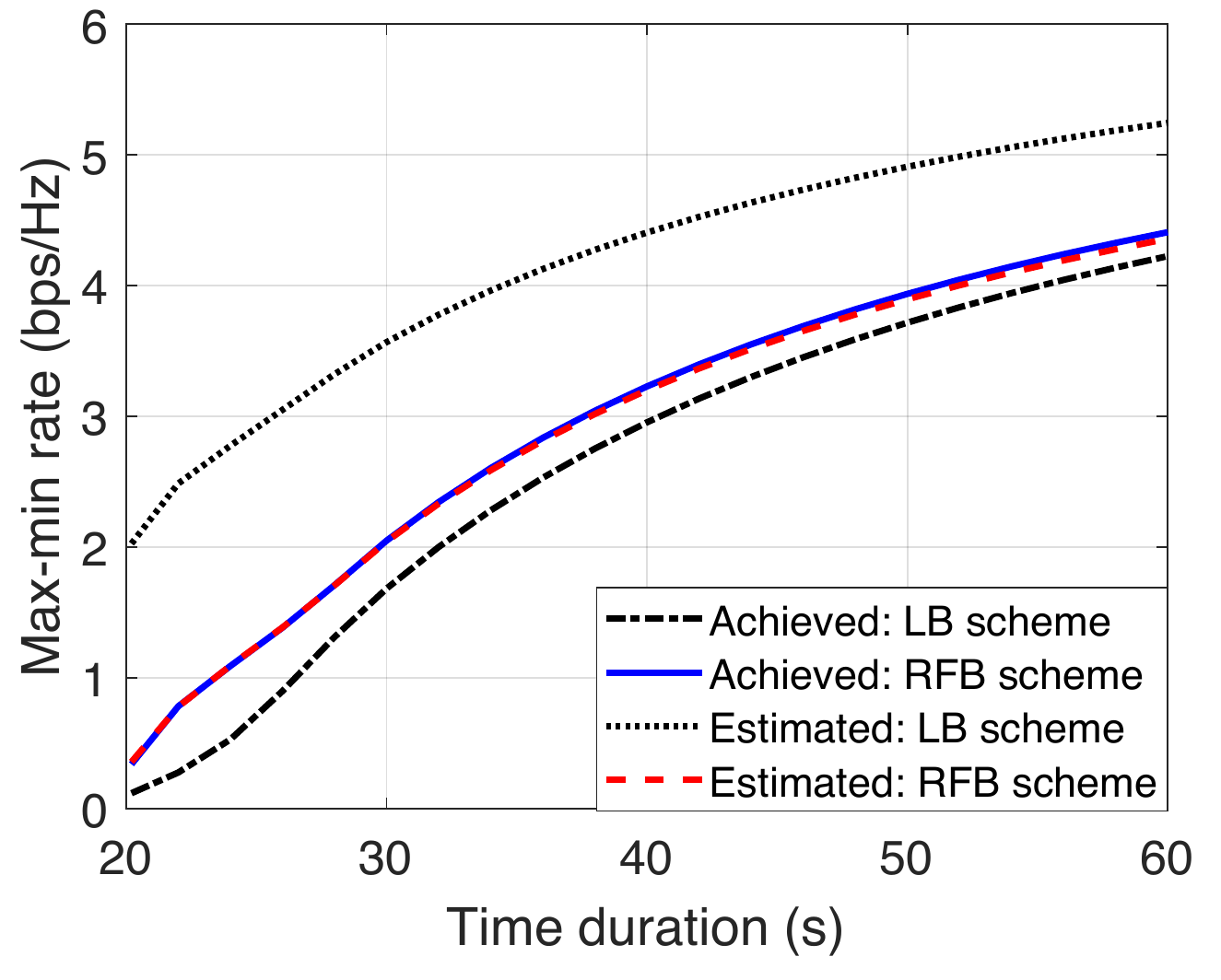}}
\hspace{10mm}
\subfigure[Instantaneous achieved rate]{\label{FigTimeInsRate}
\includegraphics[height=5.5cm]{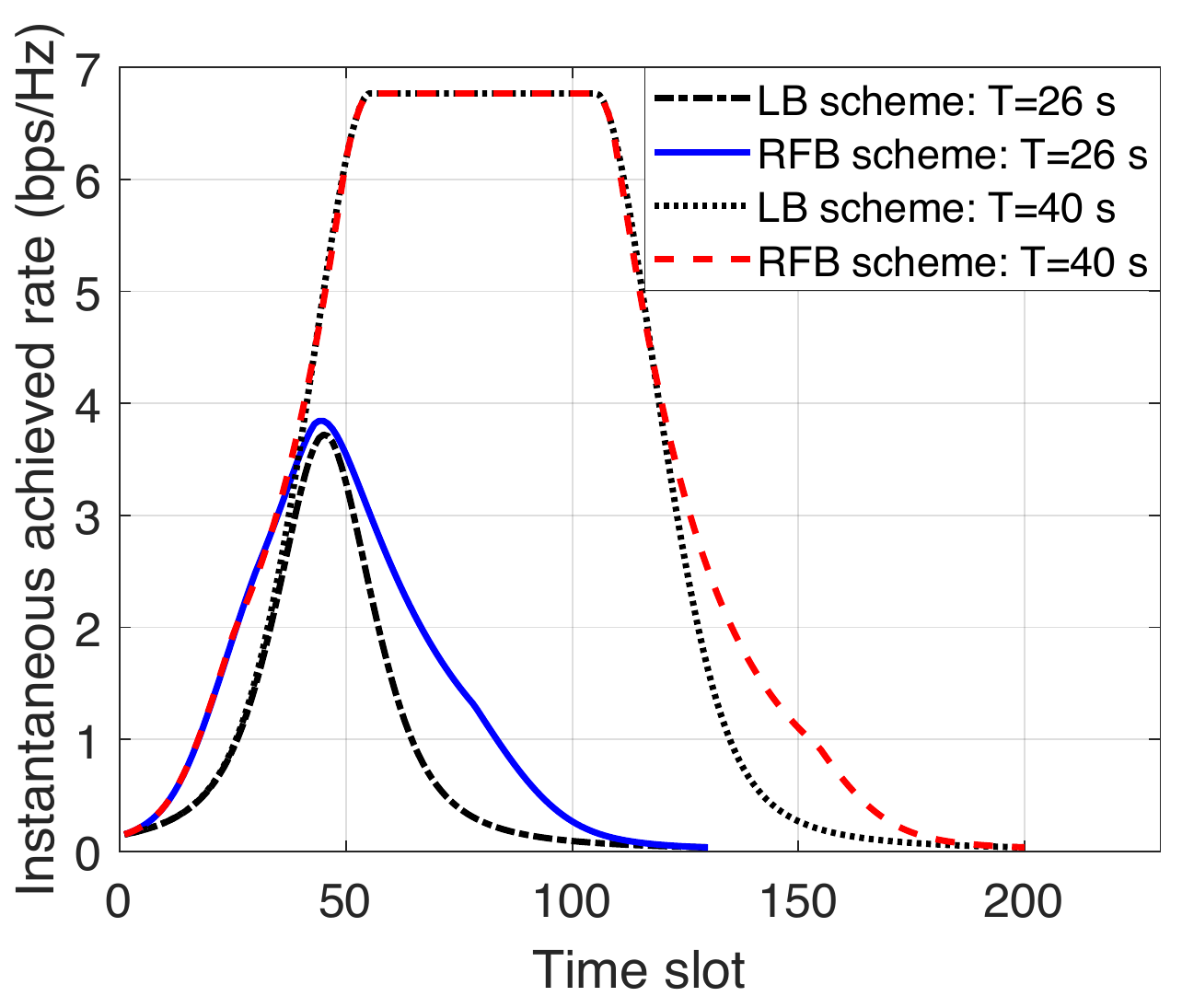}}
\caption{Comparison of the UAV trajectories and rate performance in different schemes under different time durations.}
\label{Fig:Time}
\end{figure}

Fig.~\ref{FigTimeTraj} shows the optimized UAV trajectories of LB and RFB schemes under different time durations $T$. Several interesting observations are listed as follows. First, as shown in Fig.~\ref{FigTimeTraj}, the proposed UAV horizontal trajectory is almost the same as that assuming the simplified LoS channel. This is because with one single SN, both schemes share the same principle in the horizontal trajectory design, i.e., the UAV flies  towards the SN at the maximum horizontal speed, hovers above the SN as long as possible, and then leaves to arrive at the destination in time. This leads to the elevation angle variations along the trajectory shown in Fig.~\ref{FigTimeTrajAngle}, i.e., increasing-(constant)-decreasing.  Second, compared with the LB scheme, our proposed RFB scheme can exploit an extra degrees-of-freedom (DoF) on the UAV vertical trajectory so as to adaptively adjust its altitude to balance the said distance-and-angle tradeoff. This renders the RFB scheme to achieve a better elevation angle along the trajectory, especially for the case with a short time duration (e.g., $26$ s). However, the enlargement of elevation angle diminishes when the time duration is sufficiently long (e.g., $40$ s). 
 
The effects of the time duration on the max-min rates  are shown in Fig.~\ref{FigTimeRate}, while the instantaneous achieved rates are shown in Fig.~\ref{FigTimeInsRate}.  One can observe that the estimated max-min rate of the proposed RFB scheme is close to the achieved one, while the gap is considerably large for the LB scheme. This observation validates the \emph{effectiveness} of our proposed regression method. Moreover, it indicates that it is \emph{unsuitable} to control the transmission rates using the estimated rates  obtained from the algorithm assuming the LoS channel, as it would cause unacceptable outage probability due to the ignorance of small-scale fading. Next, the achieved max-min rate of the LB scheme is only moderately less as compared to that of the RFB scheme in the regime of both short (e.g., $T$ = 20.2 s) and long (e.g., $T$ = 40 s) time durations. The underpinning reason is that in the former case, the UAV in both schemes has limited mobility DoF and thus has to almost fly straightly to the final location, leading to the similar UAV trajectories. Moreover, although the dominant instantaneous-rate regime is when the UAV flies closer to the SN,  the proposed scheme only has marginal gain in this regime due to the similar design principle as that of the LB scheme, i.e., reducing the UAV altitude when hovering above the SN.  This trend can also be observed in the case with a long time duration, for which the performance gain from the optimized vertical trajectory is diminishing since it can only marginally improve the rates in the non-dominant rate regime. Last, it is worth mentioning that the proposed RFB scheme can achieve substantial performance gain in the regime of moderate time duration, for example, it attains two folds of max-min rate when the time duration $T=26$ s.

\subsubsection{Effects of UAV maximum vertical speed}
\begin{figure}[t!]
\centering
\subfigure[3D UAV trajectory.]{\label{FigSpeedTraj}
\includegraphics[height=5.5cm]{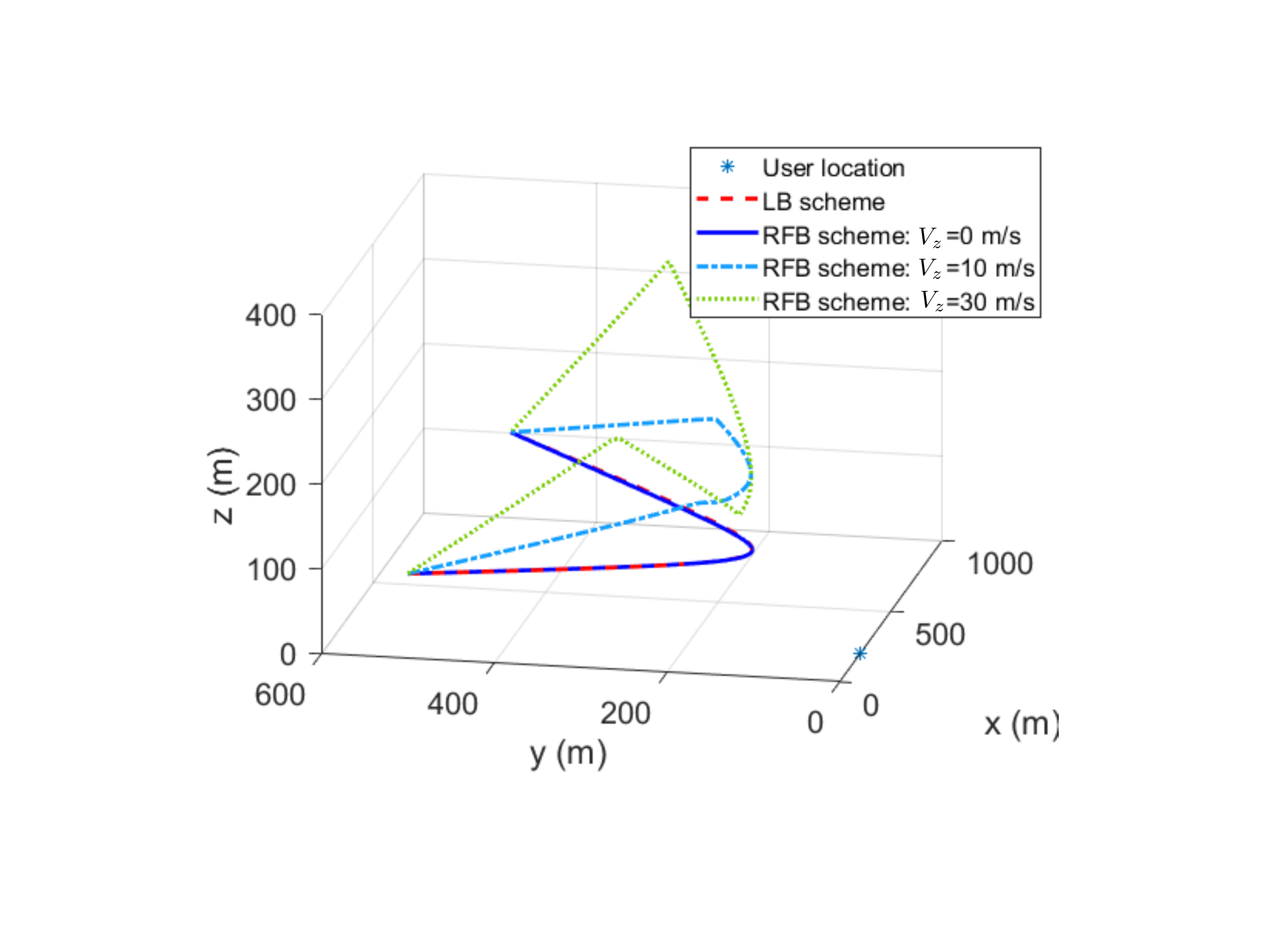}}
\hspace{10mm}
\subfigure[Max-min rate vs. UAV maximum vertical speed.]{\label{FigSpeedRate}
\includegraphics[height=5.5cm]{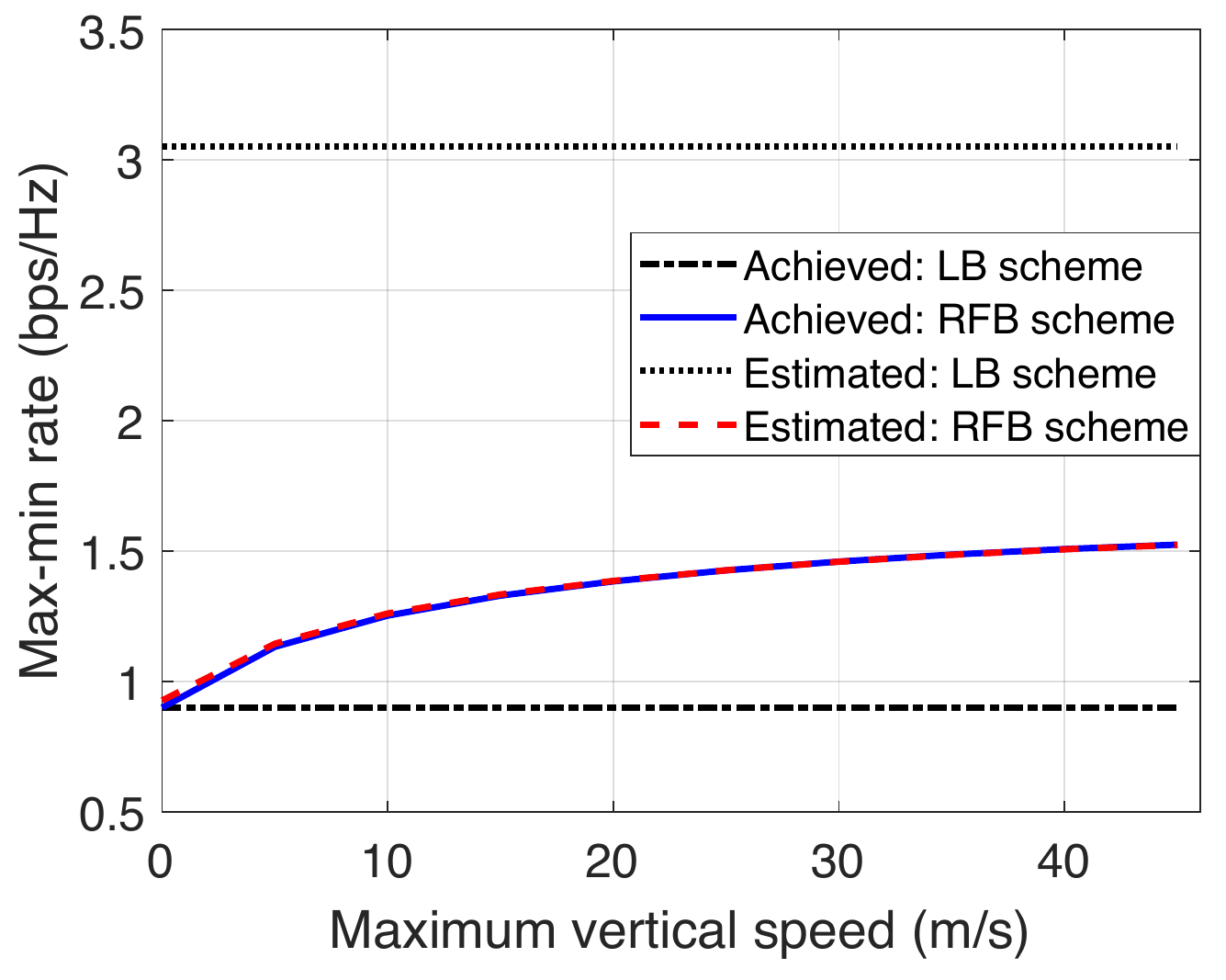}}
\caption{Comparison of the UAV trajectories and rate performance in different schemes under different maximum vertical speeds.}\label{Fig:Speed}
\end{figure}

In Fig.~\ref{FigSpeedTraj}, we plot the UAV trajectories of the LB and RFB schemes under different maximum vertical speeds. Observe that with a higher maximum vertical speed,  the UAV has more DoF to fly upwards and downwards for balancing the distance-and-angle tradeoff. The resultant performance gain can be observed in Fig.~\ref{FigSpeedRate}.  Again, the performance gap between the estimated and achieved max-min rates is negligible for the RFB scheme but significantly large  for the LB scheme. Moreover, the proposed RFB scheme has larger achieved max-min rates than the LB scheme, and the gap increases with the vertical speed and tends to saturate in the regime of high maximum vertical speed. This is expected since in that case, the UAV can fully achieve the angle gain and the vertical speed is not the performance bottleneck any more. Other observations are similar to those in Fig.~\ref{FigTimeRate}.

\subsubsection{Effects of maximum tolerable outage probability}

\begin{figure}[t!]
\centering
%\subfigure{\label{Fig:Angle_outage}
\subfigure[3D UAV trajectory.]{\label{FigOutageTraj}
\includegraphics[height=5.3cm]{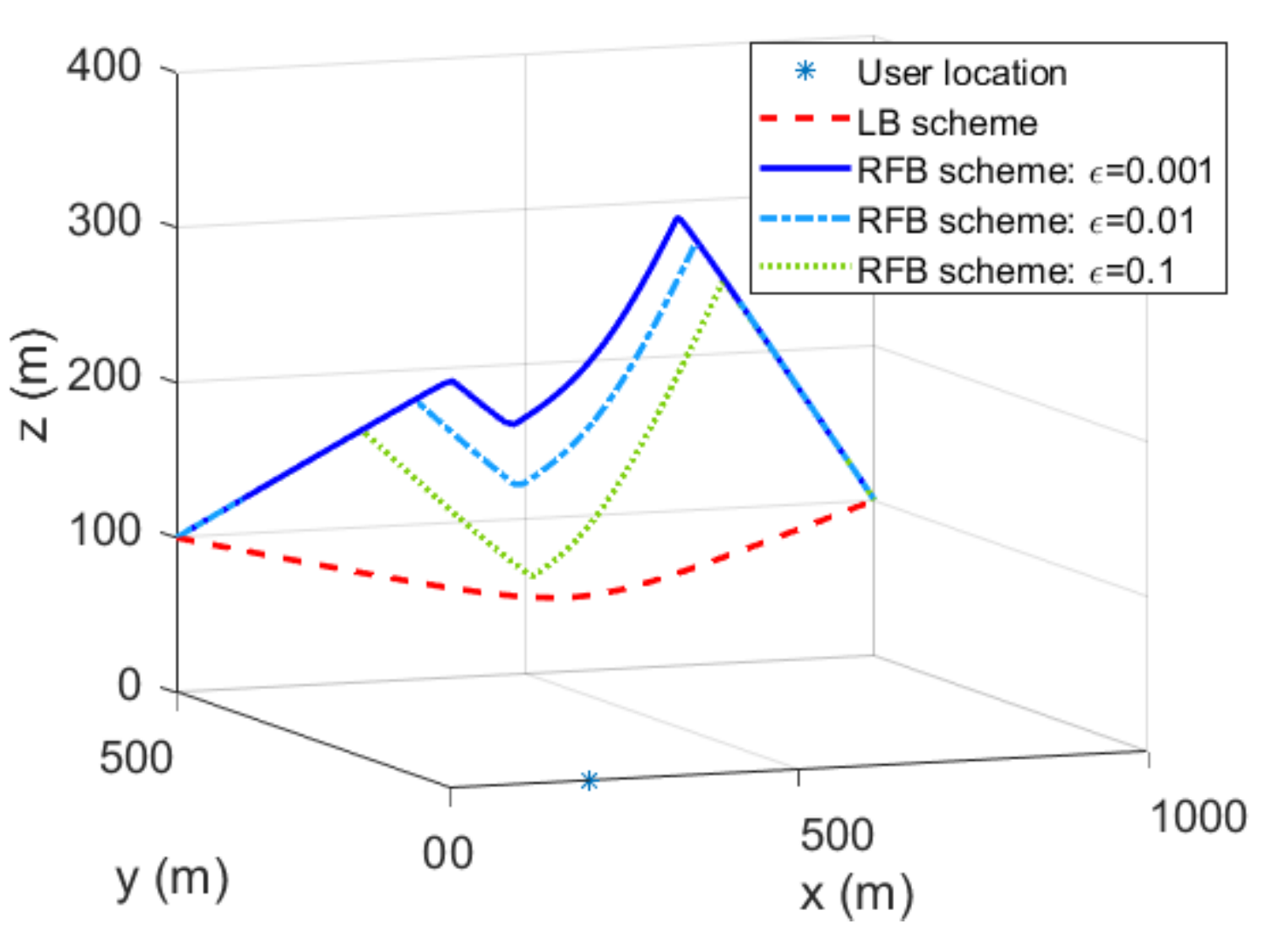}}
\hspace{10mm}
\subfigure[Max-min rate vs. maximum tolerable outage probability.]{\label{FigOutageRate}
\includegraphics[height=5.3cm]{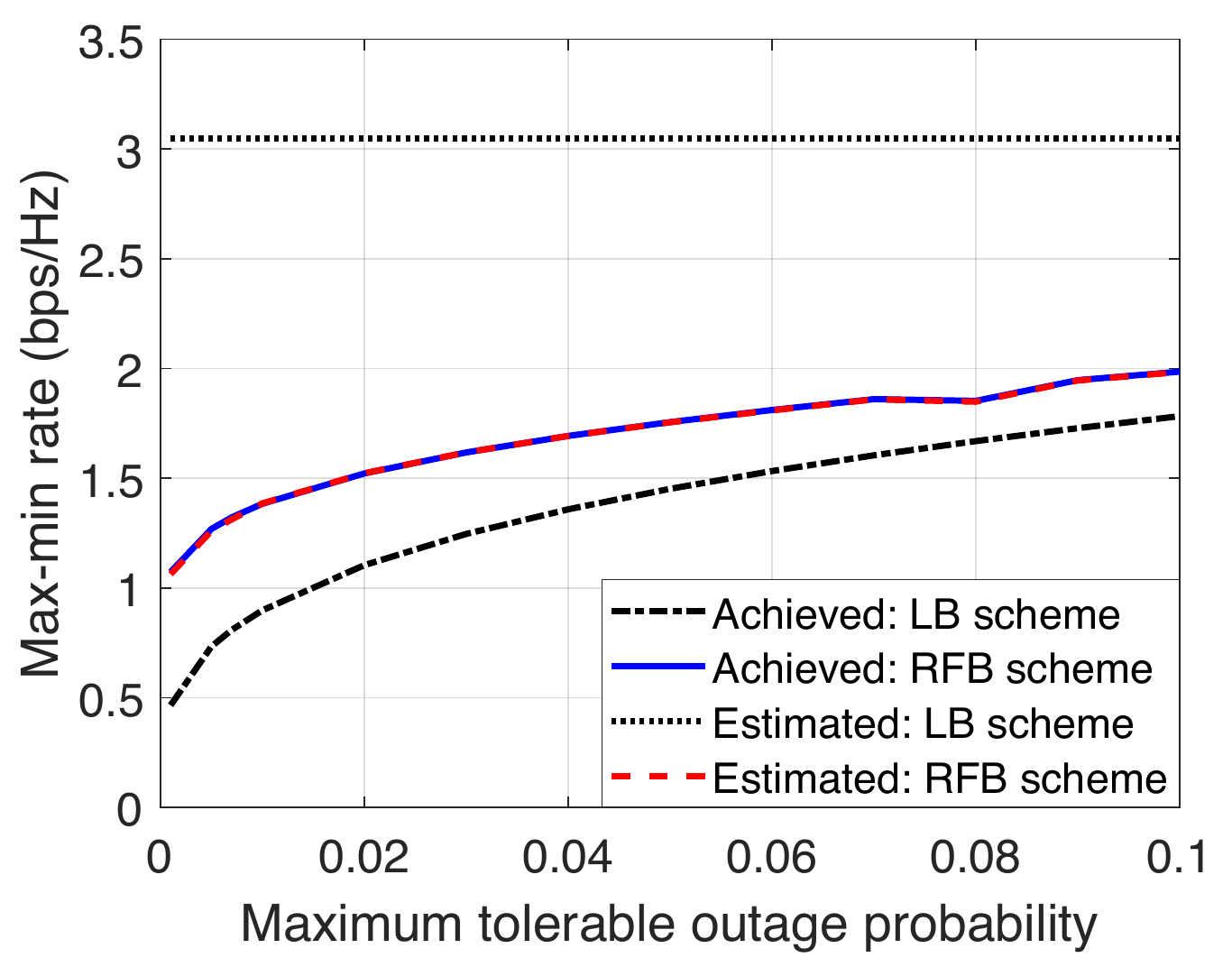}}
\caption{Comparison of the UAV trajectories and rate performance in different schemes under different maximum tolerable outage probabilities.}\label{Fig:Outage}
\end{figure} 

Fig.~\ref{FigOutageTraj} illustrates the UAV  trajectories of the LB and RFB schemes under different outage probability requirements.  It is observed that the trajectory of the LB scheme remains unchanged regardless of the value of outage probability requirement. However, for the RFB scheme, the UAV tends to fly lower given a larger maximum tolerable outage probability. The reason can be inferred from Fig.~\ref{Fig:Regression} that the angle gain is limited in this scenario since the effective fading power is already large even at a small elevation angle and the low UAV altitude incurs smaller pathloss.

The performance of max-min rate vs. the outage probability requirement is shown in Fig.~\ref{FigOutageRate}. Observe that for the case with a more stringent  outage probability requirement (e.g., for ultra-reliable communications), our proposed RFB scheme can effectively enhance the achieved max-min rate as compared to that of the LB scheme (about $1.5$ times when $\epsilon=0.01$). However, the performance gain reduces with the growth of maximum tolerable outage probability.

\subsubsection{Effects of maximum Rician factor} 
\begin{figure}[t!]
\centering
%\subfigure{\label{Fig:Angle_outage}
\subfigure[3D UAV trajectory.]{\label{FigKTraj}
\includegraphics[height=5.3cm]{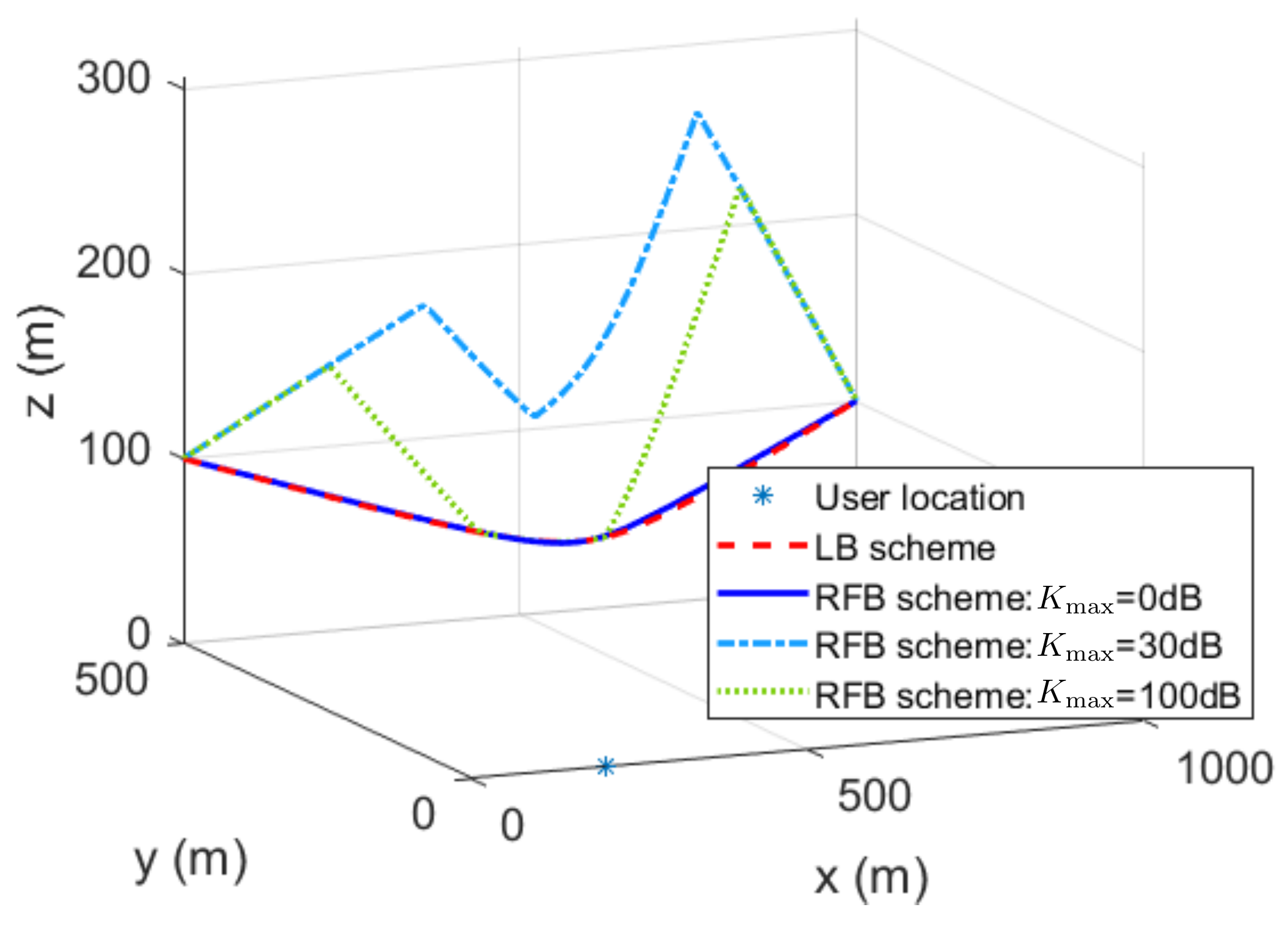}}
\hspace{10mm}
\subfigure[Max-min rate vs. maximum Rician factor.]{\label{FigKRate}
\includegraphics[height=5.3cm]{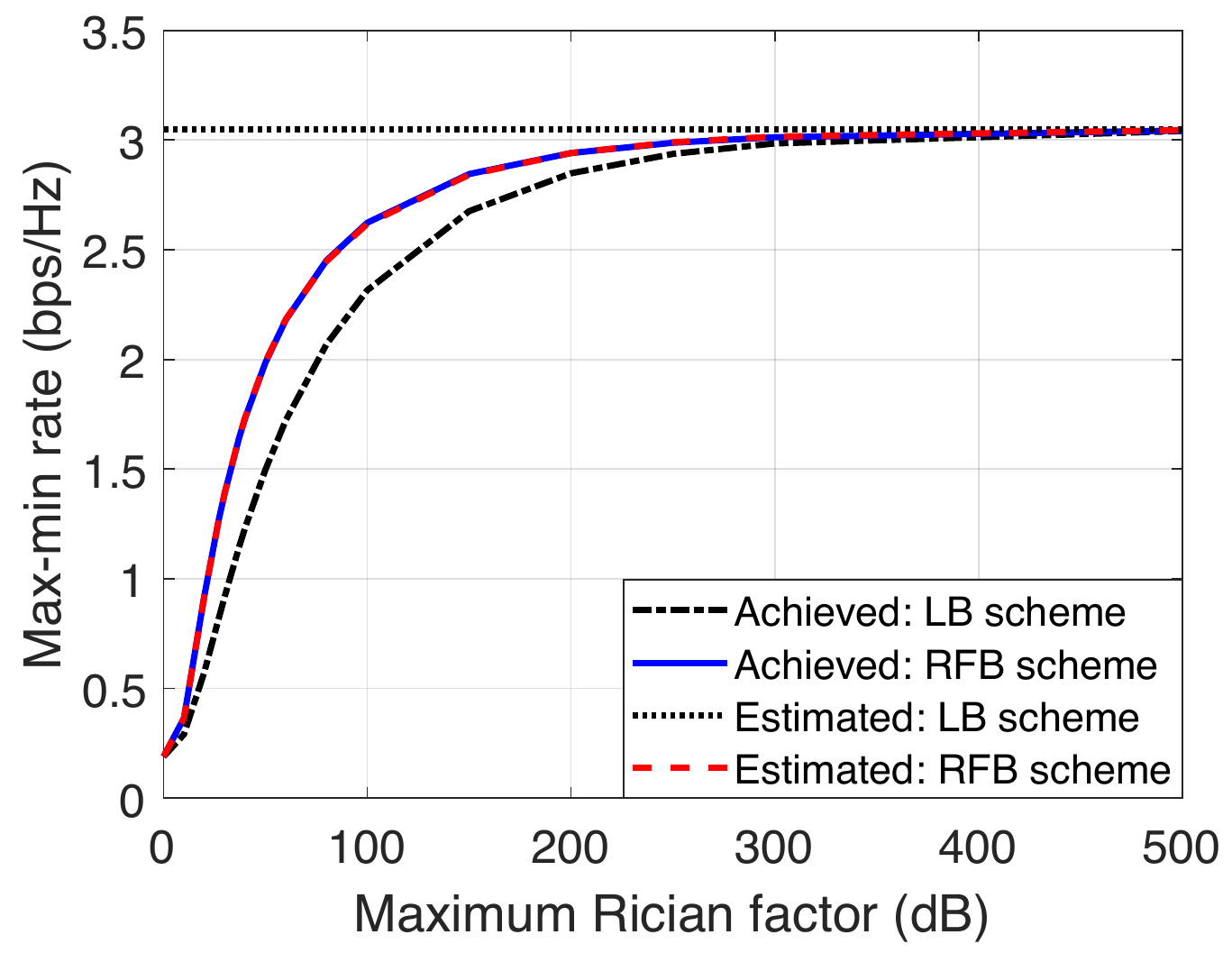}}
\caption{Comparison of the UAV trajectories and performance of max-min rate in different schemes under different maximum Rician factors.}\label{Fig:K}
\end{figure} 

The effects of the maximum Rician factor on the UAV trajectories are shown in Fig.~\ref{FigKTraj}. One interesting observation is that for the RFB scheme, with a larger maximum Rican factor, it is \emph{not} always beneficial to increase the UAV altitude. In particular, when $K_{\max}=0$ dB,  our proposed UAV trajectory reduces to that assuming the LoS channel. The reason is that the elevation angle has negligible effects on the effective fading power in this case (see Fig.~\ref{Fig:Regression}) and thus flying at the minimum altitude is optimal. With a larger maximum Rician factor, we can observe the ascent of the UAV in the overall trajectory, since a higher elevation angle can bring the considerable angle gain. However, when the maximum Rician factor is sufficiently large (e.g., $100$ dB), the proposed UAV trajectory would reduce its altitude and is expected to be equivalent to that assuming the LoS channel when $K_{\max}\to\infty$. This is because in this case, the effective fading power would dramatically increase to its maximum value at a small elevation angle and stay unchanged even further increasing the angle. Therefore, the optimal  trajectory should also fly at the lowest altitude to attain the favorable angle gain while achieving  the minimum pathloss.

The curves of max-min rate vs. the maximum Rician factor are shown in Fig.~\ref{FigKRate}. Observe that although the proposed UAV trajectory has different trends in the regimes of small and large maximum Rician factor (see Fig.~\ref{FigKTraj}), the performance of max-min rate is monotonically increasing with the growth of the maximum Rician factor and converges to that of the estimated one assuming the LoS channel when the maximum Rician factor is sufficiently large. Moreover, the performance gain of our proposed scheme firstly increases and then decreases as the maximum Rician factor increases. This is expected since the UAV trajectories in both extreme cases, i.e., with a small and infinite $K_{\max}$, reduce to those of the LB scheme.

%\section{Simulation}
\subsection{Multiple SNs}

\begin{figure}[t!]
\centering
\subfigure[3D UAV trajectory.]{\label{FigTraj4user}
\includegraphics[height=5.6cm]{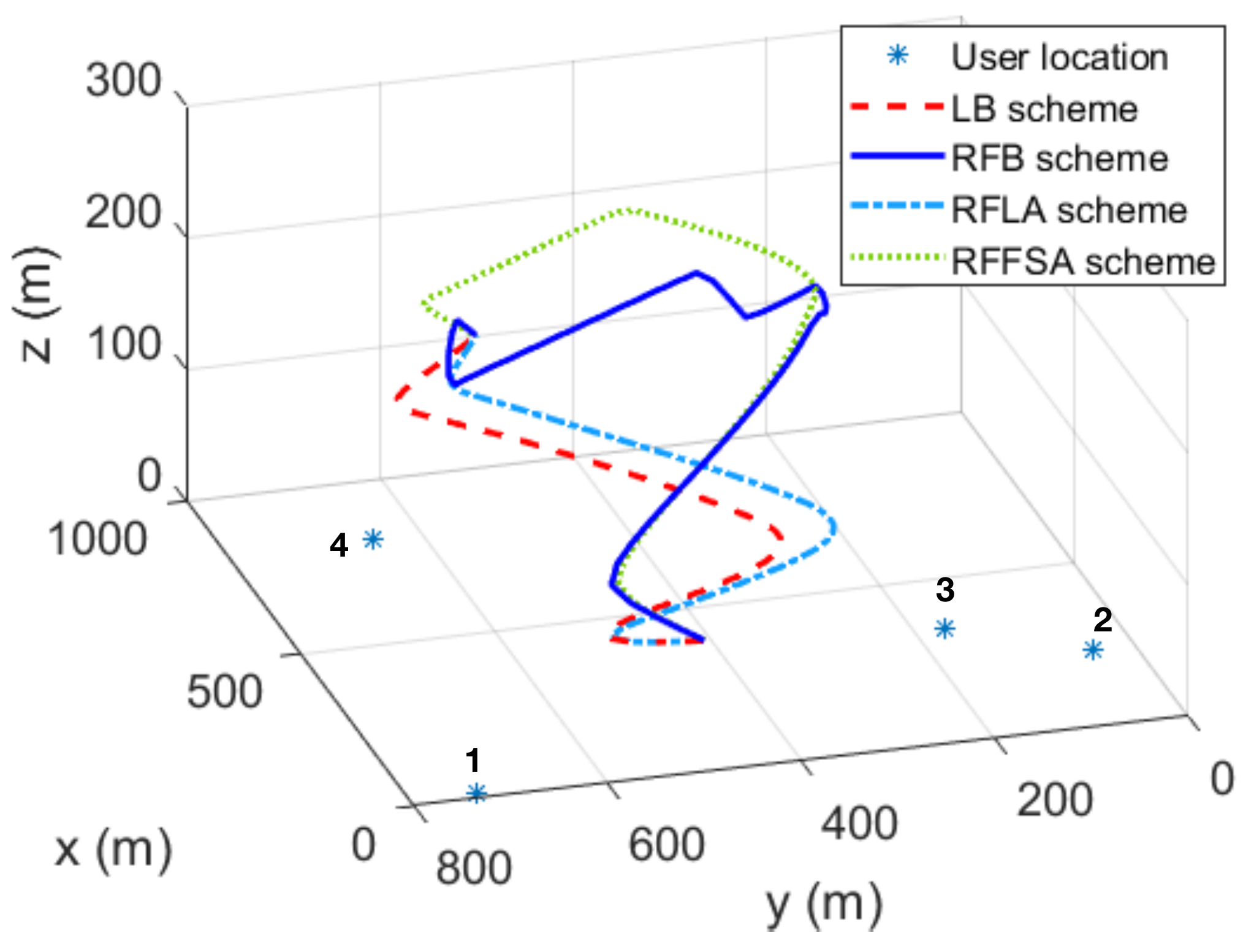}}
\hspace{10mm}
\subfigure[Horizontal trajectory.]{\label{FigTraj4userHoriz}
\includegraphics[height=5.6cm]{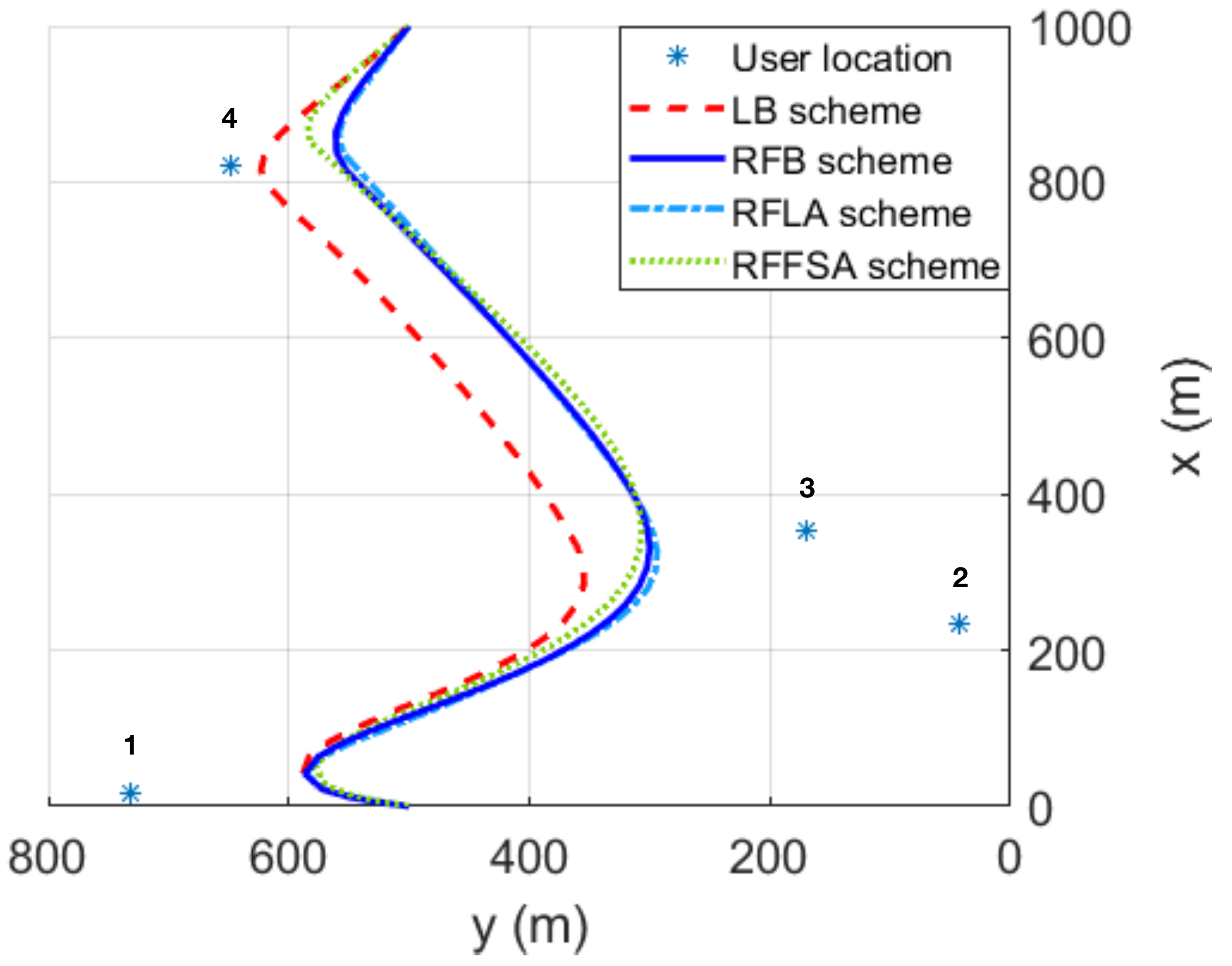}}
\subfigure[Scheduled transmission rate.]{\label{Fig4userIndRate}
\includegraphics[height=5.6cm]{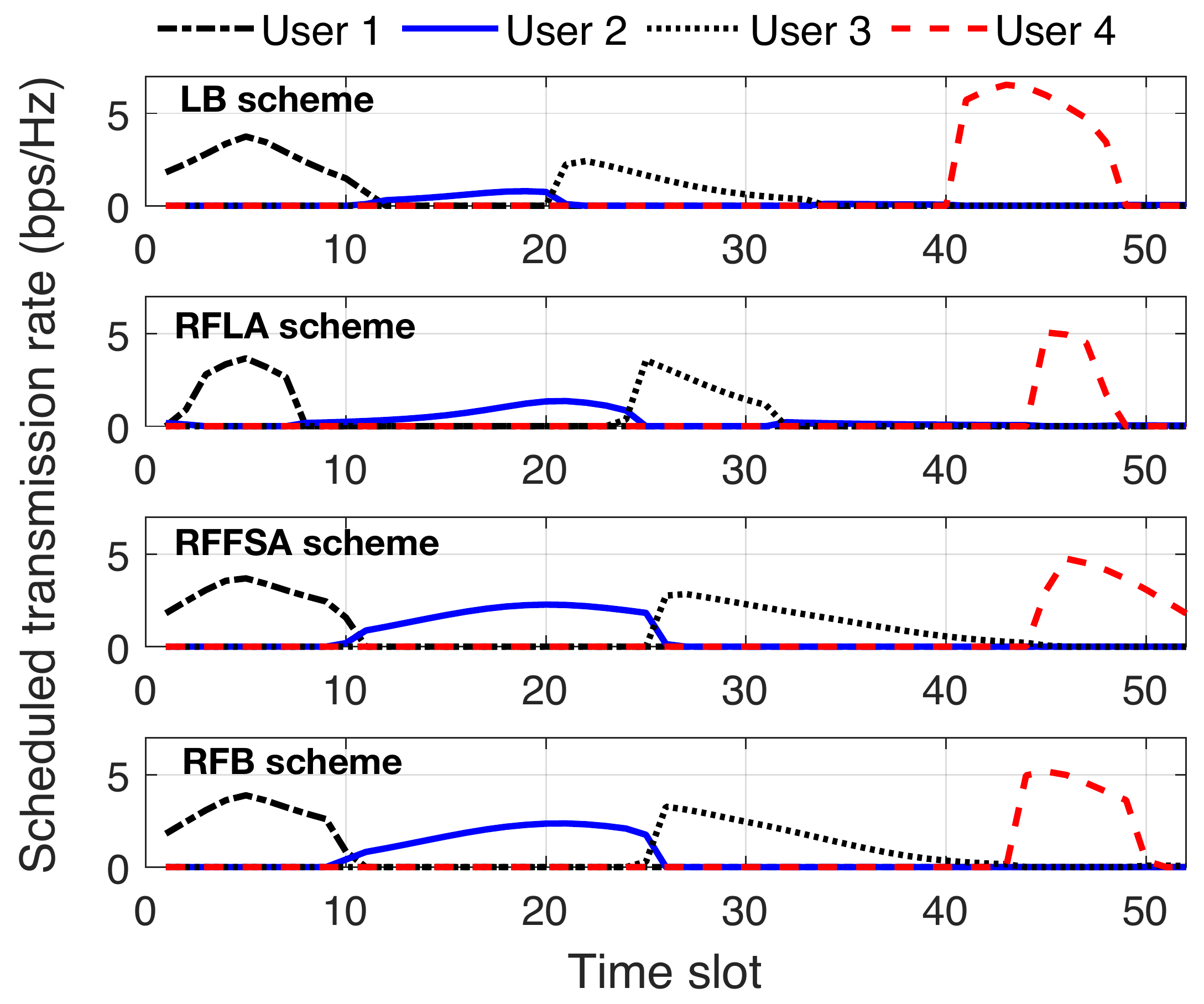}}
\hspace{10mm}
\subfigure[Max-min rate.]{\label{Fig4userRate}
\includegraphics[height=5.6cm]{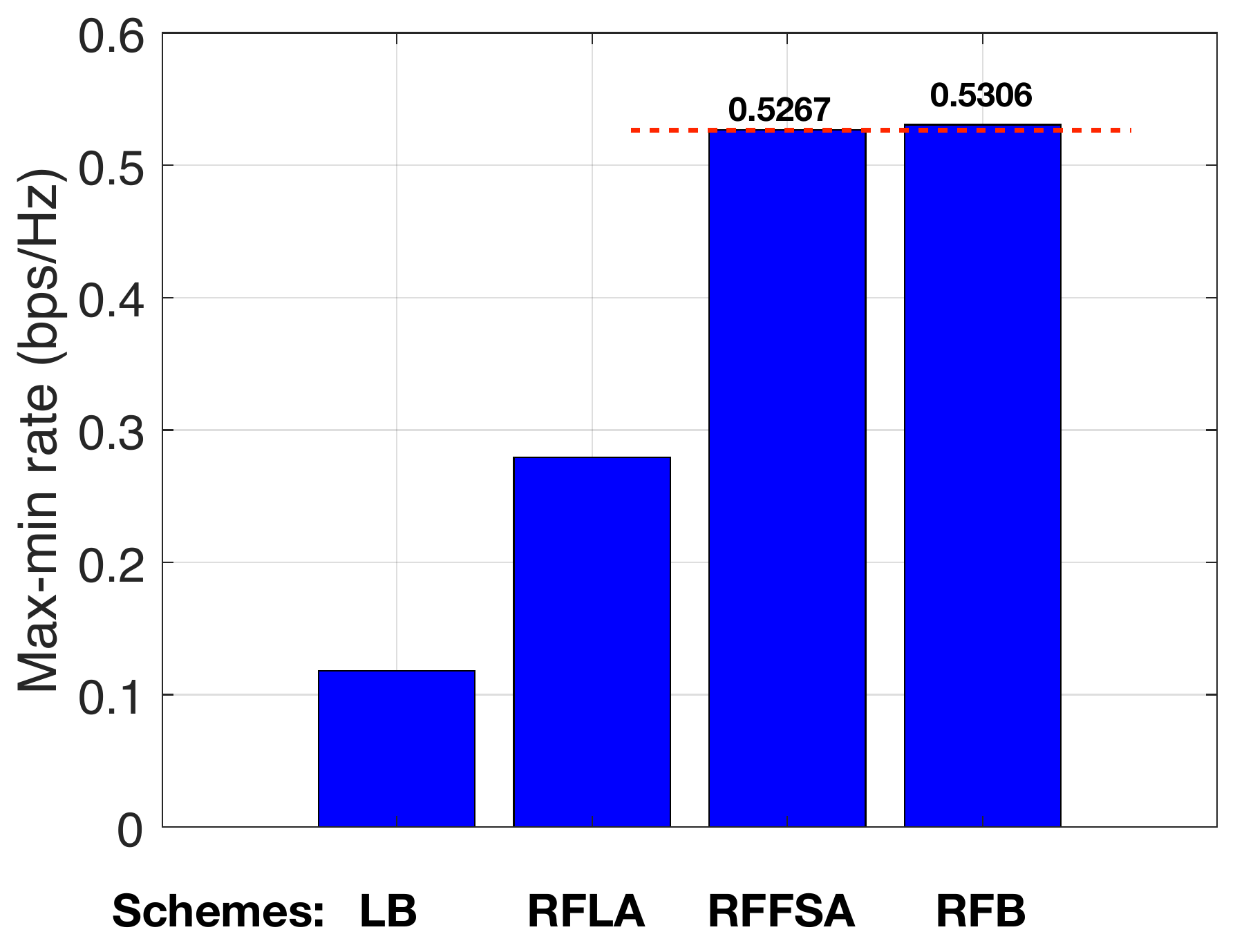}}
\caption{Comparison of the UAV trajectories and rate performance in different schemes for the case with $4$ SNs.}\label{Fig:MultiSN}
\end{figure}

Last, we consider the case with multiple SNs to evaluate the variations of designed trajectory  due to more SNs and the corresponding performance. The effects of some parameters (e.g., time duration) are similar to those of  the single-SN case and thus omitted for brevity. 
%In order to show the performance gain brought by the optimization for the vertical trajectory, in addition to the LB and RFB schemes, we consider another benchmark scheme, called Rician-fading fixed-altitude (RFFA) scheme, which only optimizes the UAV communication scheduling and horizontal trajectory proposed in this work. To avoid the complexity for the search of the optimal  altitude, we consider the same (minimum)  altitude as the LB scheme, i.e., $100$ m. 

In Figs.~\ref{FigTraj4user} and \ref{FigTraj4userHoriz}, we compare the optimized UAV trajectories by the proposed scheme with other benchmark schemes with $4$ SNs and $T=26$ s. Observe that for the horizontal trajectory, due to the limited time duration, the UAVs in all the schemes cannot sequentially visit all the SNs and stay stationary on top of each of them, following the classic  traveling-salesman-problem (TSP) solution which is known to be optimal for a sufficiently large $T$ \cite{wu2018joint}. Therefore, they can only sequentially travel nearby  each SN (resembling the TSP solution). Unlike the case with one single SN, there exist significant differences on the horizontal trajectories of the schemes assuming different channel models. In particular, compared with the LB scheme, the UAV of proposed RFB scheme gets closer to SNs $2$ and $3$ when traveling nearby them at the cost of being more away from SNs $1$ and $4$.  
%. The UAV of the RFB scheme gets even closer to these two SNs at the cost of being more away from SNs $1$ and $4$.
 The underpinning reason can be inferred from Fig.~\ref{Fig4userIndRate}, where the scheduled transmission rate $a_n[m] R_k[m]$ is shown. 
Specifically, for the LB scheme, the individual average achievable rates of SNs $2$ and $3$ are much smaller than those of SNs $1$ and $4$ due to the designed trajectory assuming the inaccurate LoS channel, which limits the network max-min rate. In contrast, the proposed RFB scheme maintains the achievable rates for SNs $1$ and $4$ by ascending the UAV altitude, and at the same time, letting the UAV travel closer to SNs $2$ and $3$ horizontally  so as to improve their rates and hence the network max-min rate as shown in Fig.~\ref{Fig4userRate}. The RFLA scheme, although having the similar horizontal trajectory as the RFB scheme, only has marginal performance gain over the LB scheme due to the lack of vertical trajectory optimization. Another interesting observation is that the RFFSA scheme has comparable performance as the RFB scheme in this case, although at the cost of huge complexity for searching the best altitude. The reason is that besides the similar horizontal trajectory, the vertical flight of the UAV in the RFFSA scheme is also analogous to that of the RFB scheme, i.e., ascending to a desirable altitude at the largest speed, hovering at that altitude, and then descending. Moreover, for the RFFSA scheme, the multiuser gain from scheduling optimization more-or-less compensates the performance loss from the trajectory optimization.  The above results show that by leveraging the angle-aware horizontal and vertical trajectory joint design, the proposed algorithm can effectively enhance the network rate  performance by balancing the achievable rates for different SNs.

\section{Conclusions}\label{Sec:Conc}
This paper considers a UAV-enabled WSN where a UAV is despatched to collect data from multiple SNs. Our objective is to maximize the minimum average data collection rate of all SNs  under the practical UAV trajectory, communication scheduling, and reliability  constraints. For the UAV-SN channels, we consider the practical angle-dependent Rician fading model with the Rician factor determined  by the UAV-SN elevation angle. The formulated optimization problem, however,  is intractable due to the lack of a closed-form expression for the effective fading power that characterizes the achievable rate. We tackle this difficulty  by approximating its function w.r.t. the 3D UAV trajectory by a logistic model using the data regression method and thereby reformulate the problem to a tractable approximate form. To solve  the non-convex reformulated problem, we propose an efficient algorithm to obtain its suboptimal solution by using the BCD and SCA techniques, and evaluate its performance numerically as compared to the benchmark designs assuming the simplified LoS channel or 2D trajectory with a fixed altitude. This work makes the first attempt to design the 3D UAV trajectory under angle-dependent Rician fading channels. The proposed approach is general and  can be  applied  to design UAV trajectories in other wireless networks and/or under other statistical channel models. For example, in multi-UAV enabled networks, the UAV cooperation can be designed by jointly optimizing their 3D trajectories as well as communication scheduling and resource allocation  under 3D collision avoidance constraints.

\appendix
\vspace{-1mm}
\subsection{Proof of Lemma~\ref{Lem:OptRelax}}\label{App:OptRelax}
This lemma is proved by contradiction. First, consider Problem P4. We can easily derive that in the optimal solution to Problem P4, the constraints \eqref{Eq:P2RateCons} for all SNs  should be active, i.e., $\frac{1}{M}\sum_{m=1}^{M} a_n^*[m] \tilde{R}_n^*[m]= \eta^*, \forall n$. Otherwise, we can always adjust $\{a_n[m]\}$ to satisfy the equality without decreasing the objective value. Next, for the constraint \eqref{Eq:P4vcons}, we assume that in the optimal solution to Problem P4, there exists a $v_n^*[m]$ such that $v_n^*[m]<\frac{z^*[m]}{\sqrt{|| {\mathbf{q}}^*[m]-{\mathbf{w}}_n||^2+z^*[m]^2}}$. Then we can always find another $\dot{v}_n[m]$ such that $\dot{v}_n[m]=\frac{z^*[m]}{\sqrt{|| {\mathbf{q}}^*[m]-{\mathbf{w}}_n||^2+z^*[m]^2}}$. With the newly chosen $\dot{v}_n[m]$, at least one of the constraints in \eqref{Eq:P2RateCons} for a SN is inactive and thus the objective value of Problem P4 can be further improved, thus contradicting to the assumption. In summary, in the optimal solution to Problem P4, both the constraints \eqref{Eq:P2RateCons} and \eqref{Eq:P4vcons} are active. Last, using the similar contradiction method, we can prove that in the optimal solution to Problem P3, the constraints  \eqref{Eq:P2RateCons} are active for all SNs. Combining these conclusions leads to the desired result.
%\vspace{-1.5mm}
{\color{black}\subsection{Proof of Lemma~\ref{Lem:ConvXY}}\label{App:ConvXY}
Let $\xi(x,y)\overset{\triangle}{=}\ln\l(1+(C_3+\frac{C_4}{x})\frac{1}{y^{\alpha/2}}\r)$ where $C_3=C_1\gamma>0$ and $C_4=C_2\gamma>0$, then $\psi(x,y)=\xi(x,y)\log_2(e)$. We first prove the convexity of $\xi(x,y)$ by the definition of convex functions. It can be obtained that the first-order derivatives of $\xi(x,y)$ w.r.t. $x$ and $y$ are
\begin{equation}
%\frac{\partial \xi(x,y)}{\partial x}
\xi_x(x,y)=\frac{-C_4}{x(xy^{\alpha/2}+C_3x+C_4)}~~ \text{and} ~~~\xi_y(x,y)=\frac{-(\alpha/2)(C_3x+C_4)}{y(xy^{\alpha/2}+C_3x+C_4)}.
\end{equation} 
Then, the Hessian of $\xi(x,y)$ is
\begin{equation}
\bigtriangledown^2 \xi(x,y)=\l[
\begin{array}{cc}
\dfrac{C_4(2xy^{\alpha/2}+2C_3x+C_4)}{x^2(xy^{\alpha/2}+C_3x+C_4)^2}& \dfrac{(\alpha/2)C_4y^{\alpha/2-1}}{(xy^{\alpha/2}+C_3x+C_4)^2}\\
\dfrac{(\alpha/2)C_4y^{\alpha/2-1}}{(xy^{\alpha/2}+C_3x+C_4)^2} & \dfrac{(\alpha/2)(C_3x+C_4)[(1+\alpha/2)xy^{\alpha/2}+C_3x+C_4]}{y^2(xy^{\alpha/2}+C_3x+C_4)^2}
 \end{array}\r].
\end{equation}
For any $\mathbf{t}=[t_1, t_2]^T$, since $\alpha\ge 2$, we have 
\begin{align}
&\mathbf{t}^T \bigtriangledown^2 \xi(x,y) \mathbf{t}\nn\\
\ge&t_1^2\l(\dfrac{C_4(2xy^{\alpha/2}+2C_3x+C_4)}{x^2(xy^{\alpha/2}+C_3x+C_4)^2}\r)+t_2^2\l(\dfrac{(C_3x+C_4)(2xy^{\alpha/2}+C_3x+C_4)}{y^2(xy^{\alpha/2}+C_3x+C_4)^2}\r)+\dfrac{2t_1t_2C_4 y^{\alpha/2-1}}{(xy^{\alpha/2}+C_3x+C_4)^2}\nn \\
=&\frac{C_4xy^{\alpha/2}(t_2x+t_1y)^2+C_4t_1^2y^2(xy^{\alpha/2}+2C_3x+C_4)+t_2^2x^2[(2C_3x+C_4)xy^{\alpha/2}+(C_3x+C_4)^2]}{x^2y^2(xy^{\alpha/2}+C_3x+C_4)^2}\ge0,\nn
\label{Eq:Hassian}
\end{align}
for $x> 0$ and $y> 0$. Therefore, $\xi(x,y)$ is a convex function, leading to the convexity of $\psi(x,y)$.}

\subsection{Proof of Lemma~\ref{Lem:RateBound}}\label{App:RateBound}
%Since $\psi(x,y)=\log_2\l(1+\l(C_1+\frac{C_2}{x}\r)\frac{\gamma}{y^{\alpha/2}}\r)$ is a convex function w.r.t. $x\ge0$ and $y\ge0$ as proved in Lemma~\ref{Lem:ConvXY}, using the SCA technique, for any given $X$ and $Y$, we have $\psi(x,y)\ge \psi(X,Y)+\psi_x(X,Y)(x-X)+\psi_y(X,Y)(y-Y), \forall X\ge0, Y\ge0,$, where
%\begin{align*}
%\psi_x(X,Y)&=\frac{-(\log_2{e})\gamma C_2}{X(XY^{\alpha/2}+\gamma C_1X+\gamma C_2))} \\
%\tilde{\psi}_y(X,Y)&=\frac{-(\log_2{e})(\alpha/2)\gamma(C_1X+C_2)}{Y(XY^{\alpha/2}+\gamma C_1X+\gamma C_2)}.
%\end{align*}
{\color{black}   Using Lemma~\ref{Lem:ConvXY}, it can be proved that $\tilde{\psi}(x,y)=\log_2\l(1+\l(C_1+\frac{C_2}{X+x}\r)\frac{\gamma}{(Y+y)^{\alpha/2}}\r)$ is a convex function w.r.t. $x\ge -X$ and $y\ge -Y$. Then using the SCA technique, for any given $x_0$ and $y_0$, we have $\tilde{\psi}(x,y)\ge  \tilde{\psi}(x_0,y_0)+\tilde{\psi}_x(x_0,y_0)(x-x_0)+\tilde{\psi}_y(x_0,y_0)(y-y_0), \forall x, y,$ where 
%\begin{equation}
%\vspace{-2mm}
%\tilde{\psi}(x,y)\ge  \tilde{\psi}(x_0,y_0)+\tilde{\psi}_x(x_0,y_0)(x-x_0)+\tilde{\psi}_y(x_0,y_0)(y-y_0), \forall x, y,
%\end{equation}
%where 
\vspace{-2mm}
\begin{align*}
\tilde{\psi}_x(x_0,y_0)&=\frac{-(\log_2{e})\gamma C_2}{(X+x_0)\l[(X+x_0)(Y+y_0)^{\alpha/2}+\gamma(C_1(X+x_0)+C_2)\r]} \\
\tilde{\psi}_y(x_0,y_0)&=\frac{-(\log_2{e})(\alpha/2)\gamma(C_1(X+x_0)+C_2)}{(Y+y_0)\l[(X+x_0)(Y+y_0)^{\alpha/2}+\gamma(C_1(X+x_0)+C_2)\r]}.
\end{align*}
By letting $x_0=0$ and $y_0=0$, we can obtain
\begin{align*}
&\log_2\l(1+\l(C_1+\frac{C_2}{X+x}\r)\frac{\gamma}{(Y+y)^{\alpha/2}}\r)\nn\\
&\ge\log_2\l(1+\l(C_1+\frac{C_2}{X}\r)\frac{\gamma}{Y^{\alpha/2}}\r)-
\frac{(\log_2{e})\gamma C_2}{X(XY^{\alpha/2}+\gamma (C_1X+C_2))}x-
\frac{(\log_2{e})(\alpha/2)\gamma(C_1X+C_2)}{Y(XY^{\alpha/2}+\gamma (C_1X+C_2))}y.
\end{align*}
By letting $\gamma=\gamma_n$, $\hat{v}_n[m]=\frac{z[m]}{\sqrt{|| {\hat{\mathbf{q}}}[m]-{\mathbf{w}}_n||^2+z[m]^2}}$, $X=1+e^{-\l(B_1+B_2 \hat{v}_n[m]\r)}$, $x=e^{-\l(B_1+B_2 v_n[m]\r)}-e^{-\l(B_1+B_2 \hat{v}_n[m]\r)}$, $Y=|| \hat{\mathbf{q}}[m]-\mathbf{w}_n||^2+z[m]^2$, and $y= || \mathbf{q}[m]-\mathbf{w}_n||^2-|| \hat{\mathbf{q}}[m]-\mathbf{w}_n||^2$, we thus derive Lemma~\ref{Lem:RateBound} where $\hat{R}_n[m]=\log_2\l(1+\l(C_1+\frac{C_2}{X}\r)\frac{\gamma}{Y^{\alpha/2}}\r)$, $\hat{\Phi}_n[m]=\frac{(\log_2{e})\gamma C_2}{X\l(XY^{\alpha/2}+\gamma(C_1X+C_2)\r)}$, and $\hat{\Psi}_n[m]=\frac{(\log_2{e})(\alpha/2)\gamma(C_1X+C_2)}{Y(XY^{\alpha/2}+\gamma (C_1X+C_2))}$.}

\vspace{-1mm}
\subsection{Proof of Lemma~\ref{Lem:vSCA}}\label{App:vSCA}
Define a function $\tilde{v}(x)=\frac{D}{\sqrt{X+x}}$. It can be easily shown that $\tilde{v}(x)$ is a convex function w.r.t. $x\ge-X$. Similar to Appendix~\ref{App:RateBound}, by using the SCA technique, for any given $x_0$, we have
\begin{equation}
\tilde{v}(x)\ge \frac{D}{\sqrt{X+x_0}}-\frac{D}{2(X+x_0)^{\frac{3}{2}}}(x-x_0).
\end{equation}
By letting $x_0=0$, we can obtain $\frac{D}{\sqrt{X+x}}\ge\frac{D}{\sqrt{X}}-\frac{Dx}{2X^{\frac{3}{2}}}.$
Last, by letting $D=z[n]$, $X=|| {\hat{\mathbf{q}}}[m]-{\mathbf{w}}_n||^2+z[m]^2$ and $x=|| {\mathbf{q}}[m]-{\mathbf{w}}_n||^2-|| {\hat{\mathbf{q}}}[m]-{\mathbf{w}}_n||^2$, we can derive Lemma~\ref{Lem:vSCA} where
\begin{align}
\hat{v}_n[m]&=\frac{z[m]}{\sqrt{|| {\hat{\mathbf{q}}}[m]-{\mathbf{w}}_n||^2+z[m]^2}}, \quad\text{and}\quad\hat{\Lambda}_n[m]&=\frac{z[m]}{2(|| {\hat{\mathbf{q}}}[m]-{\mathbf{w}}_n||^2+z[m]^2)^{\frac{3}{2}}}.
\end{align}
\vspace{-2mm}
\subsection{Proof of Convexity of Problem 11}\label{App:ConvP10}
It is observed that except the constraint \eqref{Eq:P10RateCons}, the objective function and other constraints in Problem P10 are convex. Then the remaining effort is to prove the convexity of  the constraint \eqref{Eq:P10RateCons}. To this end, we first derive the first-order derivative of $v_n[m]=\frac{z[m]}{\sqrt{|| {\mathbf{q}}[m]-{\mathbf{w}}_n||^2+z[m]^2}}$ w.r.t. $z[m]$ as follows.
\begin{equation}
\frac{\partial v_n[m]}{\partial z[m]}=\frac{|| {\mathbf{q}}[m]-{\mathbf{w}}_n||^2}{(|| {\mathbf{q}}[m]-{\mathbf{w}}_n||^2+z[m]^2)^{\frac{3}{2}}}.
\end{equation}
Then the second-order derivative of $v_n[m]$ w.r.t. $z[m]$ is
\begin{equation}
\frac{\partial^2 v_n[m]}{\partial z[m]^2}=\frac{-3z[m](|| {\mathbf{q}}[m]-{\mathbf{w}}_n||^2)}{(|| {\mathbf{q}}[m]-{\mathbf{w}}_n||^2+z[m]^2)^{\frac{5}{2}}}\le0,
\end{equation}
for $z[m]\ge H>0$. Therefore, $v_n[m]$ is concave w.r.t. $z[m]$ and thus \eqref{Eq:P10RateCons} is the convex constrain, leading to the desired result.
%\bibliographystyle{IEEEtran}
%\vspace{-4pt}
%\bibliography{BibDesk_File}
% Generated by IEEEtran.bst, version: 1.13 (2008/09/30)

\end{document}